\documentstyle[12pt]{article}

\expandafter\ifx\csname fiverm\endcsname\relax

\fi
\input prepic
\input pictex
\input postpic

\newlength{\dinwidth}
\newlength{\dinmargin}
\newfont\goth{eufm10 scaled 1200}
\newfont\bbl{msbm10 scaled 1095}
\newfont\bbs{msbm10 scaled 1000}
\newfont\bbss{msbm9 scaled 1000}
\newfont\bbsss{msbm7 scaled 1000}

\def\C{\mbox{\bbl C}}
\def\R{\mbox{\bbl R}}
\def\Z{\mbox{\bbl Z}}
\def\N{\mbox{\bbl N}}
\def\inC{\in\mbox{\bbs C}}

\def\inZ{\in\mbox{\bbs Z}}
\def\inN{\in\mbox{\bbs N}}
\def\End{{\rm End}\,}
\def\e{{\rm e}}
\def\dz{\frac{d}{dz}}
\newcommand{\Res}[2]{{\rm Res}_{#1}\left[#2\right]}
\newcommand{\D}[3]{z_#1^{-1}\delta\left(\frac{z_#2-z_#3}{z_#1}\right)}
\newcommand{\Dp}[3]{z_#1^{-1}\delta\left(\frac{z_#2+z_#3}{z_#1}\right)}
\newcommand{\Di}[3]{-z_#1^{-1}\delta\left(\frac{-z_#2+z_#3}{z_#1}
\right)}
\def\F{{\cal F}}
\def\Flie{\mbox{$\F\big/\Lt\F$\,}}
\newcommand{\Fo}[1]{\F_{(#1)}}
\newcommand{\Fn}[1]{\F^\natural_{#1}}
\def\Fger{\mbox{$\F\big/\Fo{-2}\F$\,}}
\newcommand{\Fgerq}[1]{\Fo{#1}\Big/(\Fo{-2}\F\cap\Fo{#1})}
\newcommand{\Ph}[1]{{\cal P}_{(#1)}}
\def\p{\psi}    \def\f{\varphi}
\def\V{{\cal V}}
\def\Vp{\V(\p,z)}   \def\Vf{\V(\f,z)}
\newcommand{\VP}[1]{\V(\p,z_{#1})}
\newcommand{\VF}[1]{\V(\f,z_{#1})}
\newcommand{\VV}[1]{\V(\V(\p,#1)\f,z_2)}
\def\1{{\bf 1}}      \def\w{\mbox{\boldmath$\omega$}}
\def\sig{\mbox{\boldmath$\sigma$}}
\def\bet{\mbox{\boldmath$\beta$}}
\def\Ln{{\rm L}_{(n)}}   \def\Lm{{\rm L}_{(m)}}   \def\Li{{\rm L}_{(i)}}
\def\Lmn{{\rm L}_{(m+n)}} \def\Lt{{\rm L}_{(-1)}} \def\Lo{{\rm L}_{(0)}}
\def\Le{{\rm L}_{(1)}}   \def\Lna{{\rm L}_{(-n)}}
\def\Lz{{\rm L}_{(2)}}   \def\Lza{{\rm L}_{(-2)}}
\def\vertex{$(\F,\V,\1,\w)$}
\def\lx{\left\langle}    \def\rx{\right\rangle}
\newcommand{\dual}[1]{\lx\chi^*|#1\rx}
\renewcommand{\L}{\Lambda}
\def\s{{\bf s}}   \def\r{{\bf r}}    \renewcommand{\t}{{\bf t}}
\def\Pr{\Psi_\r}   \def\Ps{\Psi_\s}
\def\alm{\alpha^\mu_m}   \def\aln{\alpha^\nu_n}   \def\al{\alpha}
\def\alf{\mbox{\boldmath$\alpha$}}
\renewcommand{\l}{\mbox{\boldmath$\lambda$}}
\def\qq{{\bf q}}   \def\pp{{\bf p}}
\def\ord{\mbox{\large\bf:}}
\def\Ord{_\times^\times}
\def\II{I\hspace{-.2em}I_{25,1}}
\def\III{I\hspace{-.2em}I_{1,1}}
\def\Leech{{\L_{\rm Leech}}}
\def\Weyl{\mbox{\boldmath$\rho$}}
\def\weyl{\rho\hspace{-.321em}\rho\hspace{-.321em}\rho}
\def\ix{\mbox{\boldmath$\xi$}}
\def\ita{\mbox{\boldmath$\eta$}}
\def\G{\mbox{\goth g}}
\def\g{\G_\F}
\def\ga{\hat{\G}(A)}
\def\cart{\hat{\mbox{\goth h}}}
\def\fake{\G_{\II}}
\def\monster{\G^{\natural\otimes\III}}
\def\su{\mbox{\goth su}}
\def\so{\mbox{\goth so}}
\newtheorem{satz}{Proposition}
\newtheorem{theo}{Theorem}
\newtheorem{defi}{Definition}

\def\beq{\begin{equation}}      \def\eeq{\end{equation}}
\def\baro{\begin{eqnarray*}}    \def\barr{\begin{eqnarray}}
\def\earo{\end{eqnarray*}}      \def\earr{\end{eqnarray}}
\newcommand{\bensatz}{\begin{satz} {\bf:} \begin{enumerate}}
\newcommand{\bsatz}{\begin{satz} {\bf:}\\ }
\newcommand{\esatz}{\end{satz}}
\newcommand{\beenden}{\end{enumerate}}
\newcommand{\beginnen}{\begin{enumerate}}
 \newcommand{\labelx}[1]{\label{#1}}
\newcommand{\BF}[1]{{\rm(}{\bf #1}{\rm)}}
\newcommand{\If}[1]{\quad\mbox{if #1}}
\newcommand{\by}[1]{\quad\mbox{by #1}}
\newcommand{\for}[1]{\quad\mbox{for #1}}
\newcommand{\lire}[2]{\displaylines{\quad{#1}\hfill\cr\hfill{}{#2}\quad
                      \cr}}
\newcommand{\rb}[1]{\raisebox{1.5ex}[-1.5ex]{#1}}
\def \ball{$\bullet$}
\def \X{0mm}
\def \Y{0mm}
\begin{document}
\thispagestyle{empty}
\setcounter{page}{0}
\renewcommand{\thefootnote}{\fnsymbol{footnote}}
\begin{flushright} DESY 93-120 \\
                   hep-th/9308151 \end{flushright}
\vspace*{2cm}
\begin{center}
{\LARGE \sc Introduction to Vertex Algebras, Borcherds Algebras, and
            the Monster Lie Algebra\footnote[1]{to appear in
            Int. J. Mod. Phys.}}\\
 \vspace*{1cm}
       {\sl Reinhold W. Gebert}\footnote[2]{Supported by
        Konrad-Adenauer-Stiftung e.V.}\\
 \vspace*{6mm}
     IInd Institute for Theoretical Physics, University of Hamburg\\
     Luruper Chaussee 149, D-22761 Hamburg, Germany\\
 \vspace*{6mm}
     August 25, 1993\\
\vspace*{1cm}
\begin{minipage}{11cm}\small
The theory of vertex algebras constitutes a mathematically rigorous
axiomatic formulation of the algebraic origins of conformal field theory
 In this context Borcherds algebras arise as certain ``physical''
subspaces of vertex algebras. The aim of this review is to give a
pedagogical introduction into this rapidly-developing area of mathemat%
ics. Based on the machinery of formal calculus we present the axiomatic
definition of vertex algebras. We discuss the connection with conformal
field theory by deriving important implications of these axioms.
In particular, many explicit calculations are presented to stress the
eminent role of the Jacobi identity axiom for vertex algebras. As a
class of concrete examples the vertex algebras associated with even
lattices are constructed and it is shown in detail how affine Lie
algebras and the fake Monster Lie algebra naturally appear. This leads
us to the abstract definition of Borcherds algebras as generalized
Kac-Moody algebras and their basic properties. Finally, the results
about the simplest generic Borcherds algebras are analysed from the
point of view of symmetry in quantum theory and the construction
of the Monster Lie algebra is sketched.
\end{minipage}
\end{center}
\renewcommand{\thefootnote}{\arabic{footnote}}
\setcounter{footnote}{0}
\newpage
\section{Introduction}
Nowadays most theoretical physicists are aware of the fact that the
present relation between mathematics and physics is characterized by
an increasing overlap between them. Prominent examples for that
development are the vertex operator construction of Kac-Moody algebras,
Calabi-Yau manifolds as possible ground states of string theory,
the sum over Riemann surfaces formulation of string theory, the
important role of modular forms and theta functions in conformal field
theory, the ``Moonshine'' meromorphic conformal field theory as a
natural representation space for the Monster sporadic simple group.
These few examples show sufficiently that it is neccessary to learn
from each other and to transfer tools and techniques.

In this work we are concerned with the mathematical theories of vertex
algebras and Borcherds algebras which were motivated and initiated by
developments in physics. After a period of rapid development which has
not ended yet we believe that it is perhaps useful to make the rigorous
mathematical framework more accessible to physicists. We shall skip the
interesting history of the subject since it is presented in great detail
in the book \cite{FLM88}. Instead we would like to emphasize that
this paper serves various purposes.

First of all we want to provide a pedagogical and self-contained
introduction into the area of vertex algebras and related subjects.
Although the reader who is familiar with conformal field theory might
devote himself to the new formalism more relaxed it could be also quite
instructive to learn vertex algebras from scratch. It is clear that in
such a treatment only the basics of the theory of vertex algebras can
be presented but after reading this review it should be no problem to
become familiar with the advanced topics in the mathematics literature.

On the other hand we hope to have compiled a comprehensive
``dictionary'' which enables a physicist to translate easily formulas
and recent results of this mathematically rigorous axiomatic formulation
of conformal field theory into his language. Throughout this paper we
will stress that vertex algebras establish a solid foundation for the
algebraic aspects of conformal field theory.

Of course, vertex algebras constitute a beautiful mathematical theory
in their own right connecting many different areas such as the
representation theory of the Virasoro algebra and affine Lie algebras,
the theory of Riemann surfaces, knot invariants, quantum groups. But
we find it especially intriguing for a physicist to see how far one
can get by purely formal manipulations as soon as the algebraic
structure is extracted from a physical theory.

We also review the present knowledge about Borcherds algebras since
up to now physicists have shown surprisingly little interest in this
topic though Borcherds algebras should be regarded as the most natural
generalization of Kac-Moody algebras. Therefore it is a reasonable
hope that they will emerge as new large symmetry algebras in
physical models. In this context especially the fake Monster Lie algebra
seemingly plays an important role in bosonic string theory.

Finally we have also included some new material. In Section 3.6
we give a natural definition of normal ordered product for vertex
operators to interpret the finiteness condition for vertex operator
algebras and to exhibit the occurrence of a Gerstenhaber-like algebraic
structure. In Section 4.6 we work out in detail the generators,
the bracket relations and the Cartan matrix for the fake Monster Lie
algebra.

Let us briefly summarize how the paper is organized.

In Section 2 we present the whole machinery of formal calculus which is
the cornerstone of the formalism. Many results are similar to those
obtained in complex analysis but here we deal solely with formal
variables and formal power series. We have collected all neccessary
formulas occuring in the relevant literature (especially \cite{FLM88}).

The first half of Section 3 is devoted to the axiomatic setup of vertex
algebras and their fundamental properties. We shall emphasize the
connection to conformal field theory. The second half is essentially
concerned with the analysis and interpretation of the Jacobi identity
for vertex algebras which should be seen as the main axiom of the
theory. As an outcome we will discuss the notions of locality and
duality, the algebra of primary fields of weight one, the cross-bracket
algebra, symmetry products. The standard reference is \cite{FrHuLe93}.

Vertex algebras associated with even lattices have their origin in
toroidal compactifications of bosonic strings. In Section 4 we construct
this important class of examples of vertex algebras (cf. \cite{Borc86}).
As an easy application we demonstrate how affine Lie algebras arise in
this context. Furthermore, the fake Monster Lie algebra \cite{Borc92}
which is the first generic example of a Borcherds algebra, is worked out
in detail.

After a definition of Borcherds algebras via generators and relations
we summarize in Section 5 the basic properties of these generalized
Kac-Moody algebras \cite{Borc91}, \cite{Borc88}. We also discuss
Slansky's investigation \cite{Slan91} of the simplest nontrivial
examples of Borcherds algebras and end with a short introduction to the
Monster Lie algebra.

In Section 6 we will finally mention the topics which we have not
treated in this introductory text and we shall give a brief status
report of the areas of current research on the field of vertex algebras.

\section{Formal Calculus}
A nice exposition of vertex operator formal calculus can be found in
\cite{FrLeMe87}. We shall closely follow \cite{FLM88} where the subject
is treated thoroughly.

\subsection{Notation}
In contrast to conformal field theory (see \cite{Bank88},\cite{Gins89}
or \cite{MooSei89}, e.g.), in the vertex algebra approach we
use {\it formal} variables $z, z_0, z_1, z_2,\dots\ $. The great
advantage of formal calculus is that we perform purely algebraic
manipulations instead of bothering about contour integrals,
single-valuedness, complex analysis etc.

The objects we will work with are formal power series. For a vector
space $W$, we set
\baro W\{z\}&=&\bigg\{\sum_{n\inC}w_nz^n|w_n\in W\bigg\} \\
      W[\![z,z^{-1}]\!]&=&\bigg\{\sum_{n\inZ}w_nz^n|w_n\in W\bigg\} \\
      W[\![z]\!]&=&\bigg\{\sum_{n\inN}w_nz^n|w_n\in W\bigg\} \\
      W[z,z^{-1}]&=&\bigg\{\sum_{n\inZ}w_nz^n|w_n\in W
                    \mbox{, almost all }w_n=0\bigg\}
                    \qquad\mbox{(Laurent polynomials)} \\
      W[z]&=&\bigg\{\sum_{n\inN}w_nz^n|w_n\in W
                    \mbox{, almost all }w_n=0\bigg\}
                    \qquad\mbox{(polynomials)} \earo
where "almost all" means "all but finitely many".\\
Note that these sets are $\C$-vector spaces under obvious pointwise
operations. We can generalize above spaces in a straightforward way
to the case of several commuting formal variables, e.g.
$W[\![z_1,z_2^{-1}]\!]=\{\sum_{m,n\inN}w_{mn}z_1^mz_2^{-n}|w_{mn}
\in W\}$. Though $W\{z\}$ may look strange at first sight due to the
sum over the complex numbers it is just another way of writing the
elements of $W^{\mbox{\bbs C}}\equiv\{f:\C\to W\}$ the space of
$W$-valued functions over $\C$.

Since we will often multiply formal series or add up an infinite number
of series it is necessary to introduce the notion of algebraic
summability.\\
Let $(x_i)_{i\in I}$ be a family in $\End W$, the vector space of
endomorphisms of $W$ ($I$ an index set). We say
that $(x_i)_{i\in I}$ is {\bf summable} if for every $w\in W$, $x_iw=0$
for all but a finite number of $i\in I$. Then the operator $\sum_{i\in I
}x_i$ is well-defined. In general an algebraic limit or a product of
formal series is defined if and only if the coefficient of {\it every}
monomial in the formal variables in the formal expression is summable.

An example of a nonexistent product is
$(\sum_{n\inN}z^n)(\sum_{m\inN}z^{-m})$
where even the coefficient of any monomial $z^l$ is not summable
(because it would be $\N$ times the identity $1\equiv{\rm id}_W$).

\subsection{$\delta$-series}
Recall that for $x\in\C$
\[ (1-x)^{-1}= \left\{ \begin{array}{c@{\quad}l} \displaystyle
             \sum_{k\inN}x^k & \mbox{if $|x|<1$}\\[1.5em] \displaystyle
 -x^{-1}\sum_{k\inN}x^{-k} & \mbox{if $|x|>1$} \end{array} \right.\]
If we define
\[\delta(z)=\sum_{n\inZ}z^n\quad\in\C[\![z,z^{-1}]\!] \]
then, formally, this is the Laurent expansion of the classical
$\delta$-function at $z=1$. Indeed, $\delta(z)$ enjoys the following
fundamental properties:
\bensatz
\item Let $w(z)\in W[z,z^{-1}]\ ,\ a\in\C^\times$. Then
      \beq w(z)\delta(az)=w(a^{-1})\delta(az) \labelx{1} \eeq
\item Let $X(z_1,z_2)\in(\End W)[\![z_1,z_1^{-1},z_2,z_2^{-1}]\!]$
      be such that
$\displaystyle \lim_{z_1\to z_2}X(z_1,z_2)$ exists (algebraically)
and let $a\in\C^\times$. Then
\barr X(z_1,z_2)\delta\left(a\frac{z_1}{z_2}\right)
             &=& X(a^{-1}z_2,z_2)\delta\left(a\frac{z_1}{z_2}\right)
   \nonumber\\
             &=& X(z_1,az_1)\delta\left(a\frac{z_1}{z_2}\right)
\labelx{2} \earr \beenden \labelx{p1} \esatz
Proof:\\
Write $w(z)=\sum_{n\inZ}w_nz^n\ ,\ X(z_1,z_2)=\sum_{m,n\inZ}
x_{mn}z_1^mz_2^n$, use the definition of $\delta(z)$ and shift
summation indices. \\[1em]
Note that $w(z)$ must be a Laurent polynomial to ensure existence of the
product with the $\delta$-series. For explicit calculations it is useful
to keep in mind that the substitutions in (\ref{1}) and (\ref{2})
correspond formally to $az=1$ and $az_1/z_2=1$, respectively. We want
to stress that in analogy with the theory of distributions an
expression like $\delta(z)\delta(z)$ does not exist. Moreover, in
(\ref{1}) and (\ref{2}) integral powers of $z$, $z_1$ and $z_2$ are
required so that $z^{1/2}\delta(z)\neq1^{1/2}\delta(z)$,
$z_1^{1/2}z_2^{1/2}\delta(\frac{z_1}{z_2})
      \neq z_2\delta(\frac{z_1}{z_2})$, e.g..

Since we can always (formally) differentiate formal power series it is
interesting to study the properties of higher derivations of
$\delta(z)$. For this purpose we consider a generating function for
all the higher drivatives,
\beq \delta(z+z_0)\equiv\e^{z_0\dz}\delta(z)=\sum_{n\inN}\frac{1}{n!}
z_0^n\delta^{(n)}(z) \labelx{3} \eeq
where $(z+z_0)^n$, $n\in\Z$, is to be expanded in nonegative powers of
$z_0$. We will come back to this convention later. It turns out that a
generalization of the formula $f(x)\delta^{(n)}(x)=(-1)^nf^{(n)}(0)
\delta^{(n)}(x)$ for the classical $\delta$-function is valid for the
formal series $\delta(z)$.
\bensatz
\item Let $p(z)\in\C[z,z^{-1}]$ and consider the derivation $D=p(z)\dz$
 of $\C[z,z^{-1}]$. Let $w(z)\in W[z,z^{-1}],\ a\in\C^\times,\ y\in
 z_0\C[\![z_0]\!]$. Then
\beq w(z)\e^{yD}\delta(az) = \left(\e^{-yD}w\right)(a^{-1})\e^{yD}
\delta(az) \labelx{4} \eeq
\item Let $p(z_1,z_2)\in\C[z_1,z_1^{-1},z_2,z_2^{-1}]$
 and consider the derivations
$D_i=p(z_1,z_2)\frac{\partial}{\partial z_i}$, $i=1,2$, of
$\C[z_1,z_1^{-1},z_2,z_2^{-1}]$. Let $a\in\C^\times$, $y\in
 z_0\C[\![z_0]\!]$ and let
$X(z_1,z_2)\in(\End W)[\![z_1,z_1^{-1},z_2,z_2^{-1}]\!]$
 be such that
$\displaystyle \lim_{z_1\to z_2}X(z_1,z_2)$ exists. Then
\barr X(z_1,z_2)\e^{yD_1}\delta\left(a\frac{z_1}{z_2}\right)
 &=& \left(\e^{-yD_1}X\right)(a^{-1}z_2,z_2)
 \e^{yD_1}\delta\left(a\frac{z_1}{z_2}\right)
   \nonumber\\
 X(z_1,z_2)\e^{yD_2}\delta\left(a\frac{z_1}{z_2}\right)
 &=& \left(\e^{-yD_2}X\right)(z_1,az_1)
 \e^{yD_2}\delta\left(a\frac{z_1}{z_2}\right)
\labelx{5} \earr \beenden \labelx{p2} \esatz
Proof:\\
Since $y$ has no constant term $\e^{yD}$ is well-defined. We have
the Leibniz rule for $D,\ w(z)\in W[z,z^{-1}],\ v(z)\in W\{z\},\
n\ge0,$
\[ D^n\big(w(z)v(z)\big)=\sum_{k=0}^{n}{n \choose k}
   \left(D^kw(z)\right)\left(D^{n-k}v(z)\right) \]
Hence
\[ \e^{yD}\big(w(z)v(z)\big)=
   \left(\e^{yD}w(z)\right)\left(\e^{yD}v(z)\right) \]
Apply $\e^{yD}$ to
$(\e^{-yD}w(z))\delta(az)=(\e^{-yD}w)(a^{-1})\delta(az)$
and invoke above formula to obtain (\ref{4}). An obvious extension of
(\ref{4}) to two variables together with (\ref{2}) gives (\ref{5}).
\\[1em]
If we read off the coefficients of $y^n$ for $n\ge0$ in above formulas
we find
\baro w(z)D^n\delta(az) &=&
 \sum_{k=0}^{n}(-1)^k{n \choose k}\left(D^kw\right)
   (a^{-1})D^{n-k}\delta(az) \nonumber\\
 X(z_1,z_2)D_1^n\delta\left(a\frac{z_1}{z_2}\right)
 &=& \sum_{k=0}^{n}(-1)^k{n \choose k}\left(D_1^kX\right)(a^{-1}z_2,z_2)
 D_1^{n-k}\delta\left(a\frac{z_1}{z_2}\right)
   \nonumber\\
 X(z_1,z_2)D_2^n\delta\left(a\frac{z_1}{z_2}\right)
 &=& \sum_{k=0}^{n}(-1)^k{n \choose k}\left(D_2^kX\right)(z_1,az_1)
 D_2^{n-k}\delta\left(a\frac{z_1}{z_2}\right) \earo
all expressions existing.

It is no restriction to consider only derivations of the form
$D=p(z)\dz,\ p(z)\in\C[z,z^{-1}]$, since the derivations of
 $\C[z,z^{-1}]$ are precisely the endomorphisms of that form.
  To see this let $d\in(\End\C)[z,z^{-1}]$ be a derivation.
 Set $p(z):=d(z)$, so that $D(z)=d(z)$. We have $D(1)=0=d(1)$
  because of $d(1)=d(1\cdot1)=d(1)+d(1)$, and $d(z^{-1})=-z^{-2}d(z)$
 because of $0=d(1)=d(z\cdot z^{-1})=d(z)z^{-1}+d(z^{-1})z$.
 This shows that $D$ and $d$ agree on all powers of $z$.
 \subsection{Expansions of zero}
 Now we want to introduce the tools for formal calculus which
correspond to contour integrals and residues for complex variables.
Define
\baro \C(z)=\C(z^{-1})&=&\{p(z)/q(z)|p(z),q(z)\in\C[z],\ q\neq0\}
                         \quad \mbox{(rational functions)} \\
      \C(\!(z)\!)&=&\{p(z)/q(z)|p(z),q(z)\in\C[\![z]\!],\ q\neq0\} \\
      \C(\!(z^{-1})\!)&=&\{p(z^{-1})/q(z^{-1})|
              p(z^{-1}),q(z^{-1})\in\C[\![z^{-1}]\!],\ q\neq0\} \earo
Elements of the latter spaces will often be expressed by analytic
functions of $z$ and $z^{-1}$, respectively. They are understood
as formal Taylor or Laurent expansions.\\
Examples:
\baro (1+z)^a &=& \sum_{n\inN}{a \choose n}z^n\in\C[\![z]\!] \\
(1+z^{-1})^a&=&\sum_{n\inN}{a \choose n}z^{-n}\in\C[\![z^{-1}]\!] \earo
In the following we will always (though sometimes not explicitly stated)
refer to the {\bf binomial convention} which says that{\it all binomial
expressions are to be expanded in nonegative integral powers of the
second variable}. This is the only point in explicit calculations at
which one must not be too sloppy. \\
Example: for $a\in\C$ the following expressions are in general not
the same
\baro \left(\frac{z_1-z_2}{z_0}\right)^a &=&
      \sum_{n\inN}{a \choose n}(-1)^nz_0^{-a}z_1^{a-n}z_2^n \\
      \left(\frac{-z_2+z_1}{z_0}\right)^a &=&
      \sum_{n\inN}{a \choose n}(-1)^{a-n}z_0^{-a}z_1^nz_2^{a-n} \earo
With the binomial convention we can rewrite the generating function for
the derivatives of the $\delta$-series as
\[ \D{0}{1}{2}=z_0^{-1}\e^{-z_2\frac\partial{\partial z_1}}\delta\left(
                   \frac{z_1}{z_0}\right) \]
As an important exercise one may prove subsequent identities which will
be extremely useful for vertex operator calculus.
\bensatz
\item \beq \D{0}{1}{2}=\Dp{1}{0}{2} \labelx{6} \eeq
\item \beq \D{0}{1}{2}\Di{0}{2}{1}=\D{2}{1}{0} \labelx{7}\eeq \noindent
where all binomial expressions are expanded in nonegative integral
powers of the second variable.
\beenden \labelx{p3} \esatz \vspace{1em}
Now let us define two canonical embeddings of the rational functions
\baro \iota_+ &:& \C(z)\hookrightarrow\C(\!(z)\!)  \\
      \iota_- &:& \C(z^{-1})\hookrightarrow\C(\!(z^{-1})\!) \earo
which denote the formal Laurent expansion in $z$ and $z^{-1}$,
respectively.\\  Example:
\baro \displaystyle \iota_+(1-z)^{-1} &=& (1-z)^{-1}=\sum_{n\inN}z^n \\
\displaystyle \iota_-(1-z)^{-1} &=& \iota_-\Big(-z^{-1}
              (1-z^{-1})^{-1}\Big)=-z^{-1}\sum_{n\inN}z^{-n} \earo
so that
\[ \delta(z)=(\iota_+-\iota_-)(1-z)^{-1} \]
We introduce a linear map $\Theta$ by
\baro \Theta\equiv\Theta_z:\C(z) &\to& \C[\![z,z^{-1}]\!] \\
             f &\mapsto& \iota_+ f-\iota_- f   \earo
We observe that ker$\Theta=\C[z,z^{-1}]$ i.e. the Laurent polynomials
are precisely those formal series for which the expansions in $z$ and
$z^{-1}$, respectively, agree. Moreover, by the partial fraction
 decomposition of a rational function, we see that the family
$\{(1-az)^{-n-1}|n\ge0,\ a\in\C^\times\}$ spans a linear complement
 of $\C[z,z^{-1}]$ in $\C(z)$.
The elements of im$\Theta$ are called {\bf expansions of zero}.
The most prominent examples of expansions of zero are given by the
$\delta$-series and its derivations.
\bsatz
For $n\in\N,\ a\in\C^\times$
\barr \frac{1}{n!}\delta^{(n)}(az) &=& \Theta\Big((1-az)^{-n-1}\Big)
\nonumber \\ &=& (1-az)^{-n-1}-(-az)^{-n-1}(1-a^{-1}z^{-1})^{-n-1}
\labelx{8} \earr i.e.
\[ \delta(z+z_0)\equiv\e^{z_0\dz}\delta(z)=
   \sum_{n\inN}\Theta\Big((1-z)^{-n-1}\Big)z_0^n  \]
\labelx{p4} \esatz
Proof:\\
The case $n=0$ being clear use the fact that $\Theta$ commutes with
 $\dz$ together with the formula
$\frac{1}{n!}\left(\dz\right)^n(1-z)^{-1}=(1-z)^{-n-1}$
to obtain the case $n>0$. \\[1em]
 Thus the set $\{\delta^{(n)}(az)|n\in\N,\ a\in\C^\times\}$
 is a basis of the space im$\Theta$ of expansions of zero.

Next we shall generalize the map $\Theta$ to the case of two formal
 variables. Let S denote the set of nonzero linear polynomials in
 variables $z_1$ and $z_2$,
\[ {\rm S}=\{az_1+bz_2|a,b\in\C,\ |a|+|b|\neq0\}\subset\C[z_1,z_2] \]
Consider the subring $\C[z_1,z_2]_{\rm S}$ of the field of rational
functions $\C(z_1,z_2)$ obtained by inverting the product of elements
 of S. We can write any $f(z_1,z_2)\in\C[z_1,z_2]_{\rm S}$
 in the form
\[ f(z_1,z_2)=\frac{g(z_1,z_2)}{z_{i_2}^s
   \prod_{l=1}^r(a_lz_{i_1}+b_lz_{i_2})} \]
where $g(z_1,z_2)\in\C[z_1,z_2],\ r,s\in\N,\ a_l\neq0\ {\rm for}\
l=1,\ldots,r$.\\
For a permutation $(i_1\;i_2)$ of $(1\;2)$ we define the map
\[ \iota_{z_{i_1}z_{i_2}}\equiv\iota_{i_1i_2}:\C[z_1,z_2]_{\rm S}\to
                                \C[\![z_1,z_1^{-1},z_2,z_2^{-1}]\!] \]
such that each factor $(a_lz_{i_1}+b_lz_{i_2})^{-1}$ in
$f\in\C[z_1,z_2]_{\rm S}$ is expanded in nonegative integral
powers of $z_{i_2}$. Clearly the maps $\iota_{i_1i_2}$ are injective.\\
To obtain "expansions of zero" in the variables $z_1$, $z_2$ we set
\baro \Theta_{i_1i_2}\equiv\Theta_{z_{i_1}z_{i_2}}:
 \C[z_1,z_2]_{\rm S}&\to& \C[\![z_1,z_1^{-1},z_2,z_2^{-1}]\!] \\
 f &\mapsto& \iota_{i_1i_2} f-\iota_{i_2i_1} f   \earo
Then ker$\Theta_{i_1i_2}=\C[z_1,z_1^{-1},z_2,z_2^{-1}]$.

We shall use the following residue notation. For a formal series
\[ w(z)=\sum_{n\inC}w_nz^n\in W\{z\} \]
we write \[\Res{z}{w(z)}=w_{-1}\]
Formal residue enjoys some properties of contour integration:
\bensatz
\item Let $\displaystyle w(z)=\sum_{n\inC}w_nz^n\in W\{z\}$.
      For $n\in\C$ \beq w_n=\Res{z}{z^{-n-1}w(z)} \labelx{9} \eeq
\item \BF{Integration by parts} Let $v(z),w(z)\in W\{z\}$. Then
      \beq \Res{z}{v(z)\dz w(z)}=-\Res{z}{w(z)\dz v(z)} \labelx{10} \eeq
\item \BF{Cauchy theorem}
      \beq \Res{z_1-z_2}{\iota_{z_1,z_1-z_2}f(z_1,z_2)}=
           \Res{z_2}{(\iota_{z_1z_2}-\iota_{z_2z_1})f(z_1,z_2)}\equiv
           \Res{z_2}{\Theta_{z_1z_2}f(z_1,z_2)} \labelx{11} \eeq
      for $\displaystyle f(z_1,z_2)=\frac{g(z_1,z_2)}
      {z_1^rz_2^s(z_1-z_2)^t}$, $r,s,t\in\Z,\
      g(z_1,z_2)\in\C[z_1,z_2]$.
\beenden \labelx{p5} \esatz
Proof:
\beginnen
\item Clear
\item Use the fact that $\Res{z}{\dz u(z)}=0$ for all $u(z)\in W\{z\}$.
\item It is sufficient to consider $h(z_1,z_2)=z_1^lz_2^m(z_1-z_2)^n,
      \ l,m,n\in\Z$.
\baro \displaystyle
\iota_{2\,1}h &=& \sum_{j\inN}(-1)^{n-j}
 {n \choose j}z_1^{l+j}z_2^{m+n-j} \\ \displaystyle
\iota_{1\,2}h &=& \sum_{j\inN}(-1)^j
  {n \choose j}z_1^{l+n-j}z_2^{m+j} \\ \displaystyle
\iota_{1,\,1-2}h &=& \sum_{j\inN}(-1)^j
 {m \choose j}z_1^{l+m-j}(z_1-z_2)^{n+j} \earo
This implies
\baro \displaystyle
\Res{z_2}{\iota_{2\,1}h} &=& (-1)^{-m-1}
 {n \choose n+m+1}z_1^{l+m+n+1} \\ \displaystyle
\Res{z_2}{\iota_{1\,2}h} &=& (-1)^{-m-1}
 {n \choose -m-1} z_1^{l+m+n+1} \\ \displaystyle
\Res{z_1-z_2}{\iota_{1,\,1-2}h} &=& (-1)^{-n-1}
 {m \choose -n-1}z_1^{l+m+n+1} \earo
i.e. we have to show
\[ -(-1)^{m+1}{n \choose n+m+1}+(-1)^{m+1}{n \choose -m-1}
   =(-1)^{n+1}{m \choose -n-1}\quad\mbox{for all }m,n\in\Z \]
\beenden
If $m,n<0$ then ${n \choose n+m+1}=0$ and above equation holds.\\
If $n+1\le 0\le m$ then ${n \choose -m-1}=0$ and above equation holds.\\
If $0\le -m-1\le n$ then ${m \choose -n-1}=0$ and above equation holds.
\\ In all other cases the binomial coefficients vanish identically.
\\[1em]
Note the wrong sign in the version of Cauchy's theorem given in \cite%
{FreZhu92} and the incorrect statement about the property of the
$\delta$-series.\\
We also mention that Cauchy's theorem is equivalent to (\ref{7}): Just
multiply (\ref{11}) for the special case $f(z_1,z_2)=z_2^m(z_1-z_2)^n$
with $z_2^{-m-1}z_0^{-n-1}$ and sum over $n,m\in\Z$ to obtain (\ref{7}).
\subsection{Projective change of variables} \labelx{2.4}
We have already used exponentials of derivatives like $\e^{z_0\dz}$
in deriving formulae for the higher derivatives of $\delta(z)$.
However, one might also expect $\e^{z_0\dz}$ to act somehow as a
one-parameter group of automorphisms (parametrized by $z_0$).
This turns out to be true in the following sense.
\bsatz
Let $\displaystyle w(z)=\sum_{m\inC}w_mz^m\in W\{z\},\ y\in z_0
\C[\![z_0]\!]$ and write $D_n=-z^{n+1}\dz,\ n\in\N$. Then we have
                                                              \beginnen
\item \BF{Translation} \beq
   \e^{-yD_{-1}}w(z)\equiv\e^{y\dz}w(z)=w(z+y) \labelx{13} \eeq
\item \BF{Scaling} \beq
   \left(\e^y\right)^{-D_0}w(z)\equiv\e^{yz\dz}w(z)=w(\e^yz)
   \labelx{14} \eeq
\item \BF{Projective change} \beq
   \e^{yD_n}w(z)=w\left((z^{-n}+ny)^{-1/n}\right)\for{}n\neq0
   \labelx{15} \eeq \labelx{p6}
\end{enumerate} with binomial convention. \esatz
Proof: \beginnen
\item Write out the expressions as sums
\item Write out the expressions as sums
\item We have $D_n=n\frac{d}{d(z^{-n})}$ for $n\neq0$.
                                                     Thus, by (\ref{13})
,\[ \e^{yD_n}\bigg(\sum_{m\inC}w_m(z^{-n})^{-m/n}\bigg)=
  \sum_{m\inC}w_m(z^{-n}+ny)^{-m/n}=w\Big((z^{-n}+ny)^{-1/n}\Big) \]
\beenden \vspace{1em} \noindent
Note that we have already made use of (\ref{13}) symbolically in
(\ref{3}) and Proposition \ref{p4}.

For later discussion of meromorphic conformal field theory (see also
\cite{Godd89b}) it is important to observe that $
                           \{D_{-1},D_0,D_1\}
                                             $ generate a representation
 of the group of M\"obius transformations by
\[ z\mapsto\frac{az+b}{cz+d},\qquad
   \left( \begin{array}{rr} a & b \\ c & d \end{array} \right)
 =\e^{\frac{b}{d}\rho(D_{-1})}d^{-2\rho(D_0)}\e^{-\frac{c}{d}\rho(D_1)},
  \qquad ad-bc=1 \]
where the identification is given by
\[ \rho:{\rm span}\{D_{-1},D_0,D_1\}\stackrel{\cong}{\longrightarrow}
         \su(1,1) \]
\[ D_{-1} \mapsto
\left( \begin{array}{rr} 0 & 1 \\ 0 & 0 \end{array} \right)
,\qquad D_0 \mapsto
\left( \begin{array}{rr} {\textstyle \frac12 } & 0 \\
0 & -{\textstyle \frac12 } \end{array} \right)
,\qquad D_1 \mapsto
\left( \begin{array}{rr} 0 & 0 \\ -1 & 0 \end{array} \right) \]
so that
\[ \e^{z_0\rho(D_{-1})}=
\left( \begin{array}{cc} 1 & z_0 \\ 0 & 1 \end{array} \right)
,\qquad \e^{z_0\rho(D_0)}=
\left( \begin{array}{cc} \e^{z_0/2} & 0 \\
0 & \e^{-z_0/2} \end{array} \right)
,\qquad \e^{z_0\rho(D_1)}=
\left( \begin{array}{cc} 1 & 0 \\ -z_0 & 1 \end{array} \right)\]

The full set of $D_n$'s, however, establishes a representation of the
{\bf Witt algebra},
\[ [D_m,D_n]=(m-n)D_{m+n} \]
the central extension of which is the essential ingredient of
two-dimensional conformal field theory.

\section{Vertex algebras}
\subsection{Axiomatics of vertex algebras} \label{3.1}
We shall give a definition of vertex (operator) algebra (cf. \cite{FrHu%
Le93}) using the notation of \cite{Godd89b} which we believe is more
accessible to physicists.
\begin{defi} {\bf:}\\
A {\bf vertex algebra} is a $\Z$-graded vector space
\[ \F=\bigoplus_{n\inZ}\Fo{n} \]
equipped with a linear map $\V:\F\to(\End\F)[\![z,z^{-1}]\!]$ which
                                                                assigns
 to each state $\p\in\F$ a {\bf vertex operator} $\Vp$, and the vertex
 operators satisfy the following axioms:
\beginnen
\item \BF{Regularity} If $\p,\f\in\F$ then
\beq \Res{z}{z^n\Vp\f}=0 \for{ n sufficiently large} \labelx
{fi} \eeq and n depending on $\p$ and $\f$
\item \BF{Vacuum} There is a preferred state $\1\in\F$, called the
vacuum, satisfying
\beq \V(\1,z)={\rm id}_\F \labelx{va} \eeq
\item \BF{Injectivity} There is a one-to-one correspondence between
states and vertex operators,
\beq \Vp=0\quad\iff\quad\p=0 \labelx{in} \eeq
\item \BF{Conformal vector}
There is a preferred state $\w\in\F$, called the
conformal vector, such that its vertex operator
\beq \V(\w,z)=\sum_{n\inZ}\Ln z^{-n-2} \labelx{cv} \eeq  \beginnen
\item gives the {\bf Virasoro algebra} with some central charge $c\in\C$
  ,\beq [\Lm,\Ln]=(m-n)\Lmn+\frac c{12}(m^3-m)\delta_{m+n,0} \labelx{vi}
  \eeq \item provides a {\bf translation generator}, $\Lt$,
 \beq \V(\Lt\p,z)=\dz\Vp \for{ every }\p\in\F \labelx{tr}\eeq
  \item gives the grading of $\F$ via the eigenvalues of $\Lo$,
  \beq \Lo\p=n\p\equiv\Delta_\p\p\for{ every }
  \p\in\Fo{n},n\in\Z \labelx{we} \eeq
  the eigenvalue $\Delta_\p$ is called the {\bf (conformal) weight} of
  $\p$.
\beenden
\item \BF{Jacobi identity} For every $\p,\f\in\F$,
\[ \lire{\D{0}{1}{2}\VP{1}\VF{2}\Di{0}{2}{1}\VF{2}\VP{1}}
{=\D{2}{1}{0}\VV{z_0}} \]
where binomial expressions have to be expanded in nonegative
integral powers of the second variable.
\beenden \labelx{d1} \end{defi}
\noindent We denote the vertex algebra just defined by \vertex.

We may think of $\F$ as the space of finite occupation number states
in a Fock space so that $\F$ is a dense subspace of the Hilbert space
$\cal H$ of states. The regularity axiom states that, given $\p,\f\in\F$
, there is always a high enough power $z^n$ such that $z^n\Vp\f$ is (at
``$z=0$'') a regular formal series. In other words, the regularity axiom
ensures that any $\Vp\f$ contains only a {\it finite} number of
singular expressions. In terms of creation and annihilation operators
it reflects the fact that any state $\f$ is killed by a finite but
large enough number of annihilation operators contained in (the normal
ordered expression) $\p_n$. We also mention that in physical
applications the vertex operator of the conformal vector corresponds
to the stress--energy tensor of the field theory.

\begin{defi} {\bf:}\\
A {\bf vertex operator algebra} is a vertex algebra with the additional
assumptions that \beginnen
\item the spectrum of $\Lo$ is bounded below
\item the eigenspaces $\Fo{n}$ of $\Lo$ are finite-dimensional. \beenden
\labelx{d2} \end{defi} \vspace{1em}
The first condition is an immediate consequence of a physical postulate.
As we will see $\Lo$ generates scale transformations. Recalling that the
variable $z$ in conformal field theory has its origin in $\e^{t+ix}$
(cf. \cite{Gins89}) one finds that $\Lo$ corresponds to time
translations. Thus it may be identified with the energy which should be
bounded below in any sensible quantum field theory. In fact, vertex
operator algebras can be regarded as a rigorous mathematical definition
of chiral algebras in physics \cite{MooSei89}. Then the formal
variable $z$ can be thought of as a local complex coordinate. The vertex
operators $\Vp$ correspond to holomorphic chiral fields i.e. they can be
viewed as operator-valued distributions on a local coordinate chart of a
Riemannn surface. In this context the three terms terms of the Jacobi
identity are geometrically interpreted as the three ways of cutting the
four--punctured Riemann sphere into two three--punctured spheres.
Alternatively, one might regard the Jacobi identity as a precise
statement of the Ward identities on the three--punctured Riemann sphere.
\cite{FreZhu92},\cite{Zhu90}

Since vertex operators are operator valued formal Laurent series we
can give an alternative formulation (see \cite{Borc92}, e.g.) of the
axioms of a vertex algebra using the mode expansion
\beq \Vp=\sum_{n\inZ}\p_nz^{-n-1} \labelx{16a} \eeq One has \beginnen
\item (Regularity)*
\beq \p_n\f=0\for{ $n$ sufficiently large} \labelx{16} \eeq
\item (Vacuum)*
\beq \1_n\p=\delta_{n+1,0}\p \labelx{17} \eeq
\item (Injectivity)*
\beq \p_n=0\quad\forall n\in\Z\quad\iff\quad\p=0 \labelx{18} \eeq
\item (Conformal vector)*
\beq \w_{n+1}=\Ln \labelx{19} \eeq
\item (Jacobi identity)*
\beq \sum_{i\ge0}(-1)^i{l \choose i}\Big(\p_{l+m-i}(\f_{n+i}\xi)-
(-1)^l\f_{l+n-i}(\p_{m+i}\xi)\Big)=\sum_{i\ge0}{m \choose i}(\p_{l+i}
\f)_{m+n-i}\xi \labelx{20} \eeq
for all $\p,\f,\xi\in\F,\ l,m,n\in\Z$.
\beenden
To see the equivalence of the two formulations of the Jacobi identity
evaluate \\ $\Res{z_2}{\Res{z_1}{\Res{z_0}{z_2^nz_1^mz_0^l
(Jacobi\ identity)}}}$. As an intermediate result one finds, by
using (\ref{6}), yet another version of the Jacobi identity which
occurs in the literature \cite{Zhu90},\cite{FreZhu92}:
\[ \lire{\Res{z_1}{\VP{1}\VF{2}\iota_{12}\left((z_1-z_2)^lz_1^m\right)-
\VF{2}\VP{1}\iota_{21}\left((z_1-z_2)^lz_1^m\right)}}
{=\Res{z_0}{\VV{z_0}\iota_{20}\left(z_0^l(z_2+z_0)^m\right)}} \]
As one might suspect the Jacobi identity contains most information of a
 vertex algebra. In fact we will see that one can derive from it
important properties such as locality and duality in conformal field
theory (cf. \cite{Godd89b}). If we put $l=m=n=0$ in (\ref{20})
we get $\p_0(\f_0\xi)-\f_0(\p_0\xi)=(\p_0\f)_0\xi$.
Later we will define an antisymmetric product on the subspace
\Flie of $\F$ by $[\p,\f]:=\p_0\f$. Then above formula
indeed establishes the classical Jacobi identity for Lie algebras.
On the other hand, choosing $\p=\f=\1$ in the Jacobi identity for vertex
algebras and using the vacuum axiom we recover (\ref{7}) and thus the
Cauchy theorem (\ref{11}). Hence the Jacobi identity for vertex algebras
may be regarded as a combination of the classical Jacobi identity for
Lie algebras and the Cauchy residue formula for meromorphic functions.

In what follows we will frequently make use of two important formulas
which are the special cases $m=0$ and $l=0$, respectively, of (\ref{20})
:\\
(Associativity formula)
\beq (\p_l\f)_n=\sum_{i\ge0}(-1)^i{l \choose i}\Big(\p_{l-i}\f_{n+i}-
                (-1)^l\f_{l+n-i}\p_i\Big) \labelx{20b} \eeq
(Commutator formula)
\beq [\p_m,\f_n]=\sum_{i\ge0}{m \choose i}(\p_i\f)_{m+n-i}
                 \labelx{20c} \eeq
for all $\p,\f\in\F,\ l,m,n\in\Z$.\\
{}From the commutator formula we can immediately infer the interesting
result that in any vertex algebra the zero mode operators, $\p_0$,
act as derivations on the products $\f_n\chi$,i.e.,
\beq \p_0(\f_n\chi)=(\p_0\f)_n\chi+\f_n(\p_0\chi) \labelx{20d} \eeq
\subsection{Basic properties of vertex algebras}
To become familiar with the definition let us derive some interesting
properties of vertex algebras (In \cite{Zhu90} some consequences of the
definition are stated incorrectly). Iterating (\ref{tr}) and using
translation (\ref{13}) we find that $\Lt$ indeed generates translations,
\beq \V\left(\e^{z_0\Lt}\p,z\right)=\V(\p,z+z_0) \labelx{27} \eeq
Moreover, the vacuum is translation invariant because (\ref{tr}) for
$\p=\1$ together with the vacuum axiom (\ref{va}) and injectivity
(\ref{in}) gives
\beq \Lt\1=0 \labelx{28} \eeq

Take $\Res{z_0}{\Res{z_1}{z_1^n(Jacobi\ identity)}}$,
\[ [\p_n,\Vf]=\sum_{i\ge0}{n \choose i}\V(\p_i\f,z)z^{n-i} \]
In the special case $\p=\w$ we obtain
\[ [\Ln,\Vf]=\sum_{i\ge-1}{n+1 \choose i+1}\V(\Li\f,z)z^{n-i} \]
in particular,
\barr [\Lt,\Vf] &=& \dz\Vf \labelx{31} \\
   \   [\Lo,\Vf] &=& \left(z\dz+\Delta_\f\right)\Vf
\If{}\f\in\Fo{\Delta_\f} \labelx{32} \earr
Using the well-known formula $\e^AB\e^{-A}=\sum_{n=0}^\infty\frac1{n!}
({\rm ad}_A)^nB\equiv\sum_{n=0}^\infty\frac1{n!}
                     \underbrace{[A,[A,\ldots[A,}_{n}B]]\ldots]$
these equations give, respectively, \\
\BF{Translation property}
\beq \e^{y\Lt}\Vf\e^{-y\Lt}=\V(\f,z+y)    \labelx{32a} \eeq
\BF{Scaling property}
\beq \e^{y\Lo}\Vf\e^{-y\Lo}=\e^{y\Delta_\f}\V(\f,\e^yz)
\If{}\f\in\Fo{\Delta_\f} \labelx{32b} \eeq
for every $y\in z_0\C[\![z_0]\!]$ by Proposition \ref{p6}.\\ Thus $\Lo$
generates scale transformations. Note that (\ref{32}) also implies
\beq \f_n\Fo{m}\subset\Fo{\Delta_\f+m-n-1}
\If{}\f\in\Fo{\Delta_\f} \labelx{33} \eeq
which means that the operator $\f_n$ shifts the grading by $\Delta_\f-
n- 1$, i.e. it can be assigned ``degree'' $\Delta_\f- n- 1$. In view
of this relation the reader might wonder again why we use subscripts
in round brackets for the grading of $\F$ and for the Virasoro
generators in contrast to the naked subscripts occuring in the mode
expansion (\ref{16a}) of a vertex operator. This possibly causes some
confusion but stems from the fact that we employ two different mode
expansions. In conformal field theory we are acquainted with the
expansion
\beq \p(z)\equiv\Vp=\sum_{n\inZ}\p_{(n)}z^{-n-\Delta_\p} \labelx{33a}
     \eeq
which depends on the conformal weight of the field $\p(z)$. To exhibit
explicitly the Virasoro algebra in the definition of a vertex algebra
we used this expansion for the vertex operator associated with the
conformal vector (stress-energy tensor!) in (\ref{cv}). It is quite easy
to convert results obtained in one expansion into the other formalism,
namely, simply by shifting the grading:
\baro \p_n&\equiv&\p_{(n+1-\Delta_\p)} \\
      \p_{(n)}&\equiv&\p_{n-1+\Delta_\p} \earo
for any homogeneous element $\p\in\F$. For example we can rewrite
(\ref{33}) as
\[ \f_{(n)}\Fo{m}\subset\Fo{m-n} \]
so that $\f_{(n)}$ always has ``degree'' $-n$ irrespective of $\f$. In
so far the mode expansion (\ref{33a}) is therefore the more natural one
because it respects the grading of $\F$. On the other hand for formal
calculus it is more useful to stick to an expansion which does not refer
to the conformal weight of a state. Hence we shall almost everywhere
in the formulae assume the mode expansion (\ref{16a}).

Let us exploit the fact that the Jacobi identity is obviously
invariant under $(\p,z_1,z_0)\leftrightarrow(\f,z_2,-z_0)$:
\baro \D{2}{1}{0}\VV{z_0}
      &=& \Dp{1}{2}{0}\V(\V(\f,-z_0)\p,z_1)\by{ symmetry} \\
 &=& \Dp{1}{2}{0}\V(\V(\f,-z_0)\p,z_2+z_0)\by{ (\ref{5})} \\
 &=& \D{2}{1}{0}\V(\V(\f,-z_0)\p,z_2+z_0)\by{ (\ref{6})} \earo
Taking $\Res{z_1}{\ldots}$ we get
\baro \VV{z_0} &=& \V(\V(\f,-z_0)\p,z_2+z_0) \\
&=&\V\left(\e^{z_0\Lt}\V(\f,-z_0)\p,z_2\right)\by{ (\ref{27})}
\earo
Injectivity (\ref{in}) finally yields \\
\BF{Skew-symmetry}
\beq \VP{0}\f=\e^{z_0\Lt}\V(\f,-z_0)\p \labelx{34} \eeq
or, in components,
\beq \p_n\f=-(-1)^n\f_n\p+\sum_{i\ge1}\frac1{i!}(-1)^{i+n+1}
     \left(\Lt\right)^i(\f_{n+i}\p) \labelx{35} \eeq

In particular, we observe that the vertex operator $\Vp$ "creates" the
state $\p\in\F$ when applied to the vacuum:
\beq \Vp\1=\e^{z\Lt}\p \labelx{36} \eeq
by (\ref{va}). In components,
\beq \p_n\1=\cases{0 &for $n\ge0$ \cr
                   \p &for $n=-1$ \cr
 \frac1{(-n-1)!}\left(\Lt\right)^{-n-1}\p &for $n\le-2$} \labelx{37}\eeq
Hence the vacuum satisfies
\beq \Ln\1=0\qquad\forall n\ge{-1} \labelx{38}\eeq

We shall denote by $\Ph{\Delta}$ the space of {\bf(conformal) highest
weight vectors {\rm or} primary states} satisfying
\barr \Lo\p &=& \Delta\p\qquad\mbox{i.e. }\p\in\Fo{\Delta} \nonumber\\
     \Ln\p &=& 0\qquad\forall n>0 \labelx{39} \earr
Thus in any vertex algebra the vacuum is a primary state of weight zero.
We immediately find for $\p\in\Ph{\Delta}$
\beq [\Ln,\Vp] = z^n\left\{z\dz+(n+1)\Delta\right\}\Vp \qquad\forall
n\in\Z \labelx{40} \eeq or \beq
[\Ln,\p_m]=\left\{(\Delta-1)(n+1)-m\right\}\p_{m+n} \qquad\forall
m,n\in\Z \labelx{41} \eeq
i.e. $\Vp$ is a so called {\bf(conformal) primary field} of weight
$\Delta$.
We can rewrite (\ref{40}) as \[
[\Ln,z^{\Delta(n+1)}\Vp]=z^{n+1}\dz\left\{z^{\Delta(n+1)}\Vp\right\} \]
so that, by (\ref{15}),
\beq \e^{y\Ln}\Vp\e^{-y\Ln}=\left(\frac{\partial z_1}{\partial z}\right)
^\Delta\VP{1}\qquad\forall n\neq0 \labelx{42} \eeq
for every $y\in z_0\C[\![z_0]\!]$ where
$z_1=(z^{-n}-ny)^{-1/n}=z(1-nyz^n)^{-1/n}$

The operators $\{\Lt,\Lo,\Le\}$ satisfy the $\su(1,1)$ Lie algebra
\[ [\Lo,\Le]=-\Le,\qquad[\Lo,\Lt]=\Lt,\qquad[\Le,\Lt]=2\Lo \]
Hence we have established the following M\"obius transformation
properties of the vertex operators (see also \cite{Godd89b}):\\
If $\p\in\F$ is a {\bf quasi-primary state} of weight $\Delta$, i.e.
$\p$ satisfies $\Ln\p=\delta_{n,0}\Delta\p,\ n=0,1,$ then
\[ D_\gamma\Vp D_\gamma^{-1}=\left(\frac{d\gamma(z)}{dz}\right)^\Delta
   \V(\p,\gamma(z)) \]
where
\[ \gamma(z)=\frac{az+b}{cz+d},\qquad
   D_\gamma=\e^{\frac bd\Lt}\left(\frac{\sqrt{ad-bc}}d\right)^{2\Lo}
            \e^{-\frac cd\Le}\for{}a,b,c,d\in z_0\C[\![z_0]\!] \]
(cf. end of Subsection \ref{2.4})

Now equation (\ref{38}) tells us that the vacuum vector is SU(1,1)
invariant and the question arises whether the state $\1$ is uniquely
(up to scalar multiples) characterized by this property. In general the
answer is no but SU(1,1) invariant states come quite close to the
properties of the vacuum. To see this suppose that $\kappa\in\F$
satisfies $\Ln\kappa=0,\ n=0,\pm1$. Then, by injectivity and translation
, $\Lt\kappa=0$ is equivalent to $\kappa_n=\delta_{n+1,0}\kappa_{-1}$
in agreement with (\ref{17}). However, as $\kappa_{-1}$ regards the
associativity formula (\ref{20b}) and the commutator formula (\ref{20c})
yield $(\kappa_{-1}\f)_n=\kappa_{-1}\f_n=\f_n\kappa_{-1}\,\forall\f\in
\F,n\in\Z$. For general vertex algebras this does {\it not} force
$\kappa_{-1}$ to be a scalar multiple of the identity but rather states
that $\kappa_{-1}$ may be regarded as a Casimir operator of the vertex
algebra. In the case of a {\it simple} vertex algebra (i.e. the vertex
algebra constitutes an irreducible module for itself, cf. \cite{FrHuLe9%
3}) we can apply Schur's lemma \cite{FulHar91} to infer that
$\kappa_{-1}$ is indeed a scalar multiple of the identity.

Finally, using the Virasoro algebra (\ref{vi}) and (\ref{37}) we find
\baro \w_l\w &=& \w_l(\w_{-1}\1) \\
             &=& [\w_l,\w_{-1}]\1+\w_{-1}(\w_l\1) \\
    &=& (l+1)\w_{l-2}\1+\frac c2\delta_{l,3}\1+\w_{-1}(\w_l\1) \earo
i.e. (cf. \cite{Borc92}) \baro \w_1\w&=&2\w \\ \w_2\w&=&0 \\ \w_3\w&=&
{\textstyle \frac c2} \\ \w_l\w&=&0\for{ }l>3 \earo
In particular, $\w$ is a quasi-primary state of conformal weight two. We
want to mention that $\w$ is characterized by above relations together
with (\ref{we}) and $\w_0\p=\p_{-2}\1$ for $\p\in\F$.
\subsection{Locality and duality for vertex algebras}
To complete the relation between vertex algebras and conformal field
theory we consider matrix elements of products of vertex operators.
Define the {\bf restricted dual} of $\F$,
\[ \F^\prime\equiv\bigoplus_{n\inZ}\Fo{n}^* \]
the direct sum of the dual spaces of the homogeneous subspaces $\Fo{n}$,
i.e. the space of linear functionals on the vertex algebra $\F$
vanishing on all but finitely many $\Fo{n}$. We shall use $\langle\_|\_
\rangle$ for the natural pairing between $\F$ and $\F^\prime$. From the
regularity axiom and (\ref{33}) it is clear that any matrix element of
a vertex operator is a Laurent polynomial in $z$,
\beq \dual{\Vp\f}\in\C[z,z^{-1}]
   \for{ all }\chi^*\in\F^\prime,\ \p,\f\in\F \labelx{45} \eeq
and in this sense these 3-point correlation functions may be regarded as
meromorphic functions of $z$. Of course, we identify formally $\chi^*$
with an ``out-state'' inserted at $z=\infty$ and $\f$ with an ``in-state
'' inserted at $z=0$.
We have the following important theorem due to \cite{FLM88}.
\begin{theo} {\bf:}\labelx{t1} \beginnen
\item \BF{Locality $\equiv$ rationality of products \& commutativity}\\
For $\chi^*\in\F^\prime,\ \p,\f,\xi\in\F$, the formal series
$\dual{\VP{1}\VF{2}\xi}$ which involves only finitely many negative
powers of $z_2$ and only finitely many positive powers of $z_1$, lies
in the image of the map $\iota_{12}$:
\beq \dual{\VP{1}\VF{2}\xi}=\iota_{12}f(z_1,z_2) \labelx{46} \eeq
where the (uniquely determined) element $f\in\C[z_1,z_2]_{\rm S}$
is of the form
\[ f(z_1,z_2)=\frac{g(z_1,z_2)}{z_1^rz_2^s(z_1-z_2)^t} \]
for some polynomial $g(z_1,z_2)\in\C[z_1,z_2]$ and $r,s,t\in\Z$.\\
We also have
\beq \dual{\VF{2}\VP{1}\xi}=\iota_{21}f(z_1,z_2) \labelx{47} \eeq
i.e. $\VP{1}\VF{2}$ agrees with $\VF{2}\VP{1}$ as operator-valued
rational functions.
\item \BF{Duality  $\equiv$ rationality of iterates \& associativity}\\
For $\chi^*\in\F^\prime,\ \p,\f,\xi\in\F$, the formal series
$\dual{\VV{z_0}\xi}$ which involves only finitely many negative powers
of $z_0$ and only finitely many positive powers of $z_2$, lies in the
image of the map $\iota_{20}$:
\beq \dual{\VV{z_0}\xi}=\iota_{20}f(z_0+z_2,z_2) \labelx{48} \eeq
with the same f as above,
\beq \dual{\V(\p,z_0+z_2)\VF{2}\xi}=\iota_{02}f(z_0+z_2,z_2)
\labelx{49} \eeq
i.e. $\VP{1}\VF{2}$ agrees with $\VV{z_1-z_2}$
as operator-valued rational functions, where the right-hand expression
is to be expanded as a Laurent series in $z_1-z_2$.
\beenden \end{theo}
Proof (taken from \cite{FLM88},\cite{FrHuLe93}): \beginnen
\item Using (\ref{7}) we can rewrite the Jacobi identity in the form
\[ \lire{\D{0}{1}{2}\VP{1}\VF{2}\Di{0}{2}{1}\VF{2}\VP{1}}
   {=\V\left(\left(\D{0}{1}{2}\V(\p,z_1-z_2)
   \Di{0}{2}{1}\V(\p,-z_2+z_1)\right)\f,z_2\right)} \]
Taking $\Res{z_0}{\ldots}$ leads to the commutator formula
\[ [\VP{1},\VF{2}]=
    \V\left(\left(\V(\p,z_1-z_2)-\V(\p,-z_2+z_1)\right)\f,z_2\right) \]
By (\ref{45}), the matrix element $\dual{\ldots\xi}$ of the right-hand
side is clearly an expansion of zero in the variables
$z_1,\ z_2$ of the following form:
\[ \dual{\V\left(\left(\V(\p,z_1-z_2)-\V(\p,-z_2+z_1)\right)
   \f,z_2\right)\xi}=\Theta
    _{12}\left(\frac{g(z_1,z_2)}{z_2^s(z_1-z_2)^t}\right) \]
with $g(z_1,z_2),s,t$ as stated above. Thus
\[ \dual{\VP{1}\VF{2}\xi}-\iota
    _{12}\left(\frac{g(z_1,z_2)}{z_2^s(z_1-z_2)^t}\right)
   =\dual{\VF{2}\VP{1}\xi}-\iota
    _{21}\left(\frac{g(z_1,z_2)}{z_2^s(z_1-z_2)^t}\right) \]
But the left-hand side involves only finitely many positive powers of
$z_1$, by (\ref{33}), and the right-hand side involves only finitely
many negative powers of $z_1$, by the regularity axiom. If we further
take into account that, by (\ref{45}), the coefficient of each power
of $z_1$ on either side is a Laurent polynomial in $z_2$ then
\[ f(z_1,z_2):=\frac{g(z_1,z_2)}{z_2^s(z_1-z_2)^t}
   +{\rm h}(z_1,z_2) \]
for some h$(z_1,z_2)\in\C[z_1,z_1^{-1},z_2,z_2^{-1}]$ satisfies the
desired conditions.
\item Using (\ref{6}) and (\ref{5}) we can rewrite the Jacobi identity
    in the form
\[ \lire{\Dp{1}{0}{2}\V(\p,z_0+z_2)\VF{2}\Di{0}{2}{1}\VF{2}\VP{1}}
{=\Dp{1}{2}{0}\VV{z_0}} \]
Thus \baro
\lefteqn{\Dp{1}{2}{0}\VV{z_0}-\Dp{1}{0}{2}\V(\p,z_0+z_2)\VF{2}} \\
&=&\hspace{-.18em}\left\{\D{2}{1}{0}-\D{0}{1}{2}\right\}\VF{2}\VP{1}
                                                     \by{(\ref{7})} \\
&=&\hspace{-.18em}\VF{2}\left\{\Dp{1}{2}{0}\V(\p,z_2+z_0)-\Dp{1}{0}{2}
          \V(\p,z_0+z_2)\right\}\quad\mbox{by }(\ref{6}),(\ref{5}) \earo
Taking $\Res{z_1}{\ldots}$ leads to
\[ \VV{z_0}-\V(\p,z_0+z_2)\VF{2}
    =\VF{2}\left\{\V(\p,z_2+z_0)-\V(\p,z_0+z_2)\right\} \]
We use this formula in place of the commutator formula and
apply the same arguments as in part one to obtain the desired result.
To get the last statement put formally $z_0=z_1-z_2$.
\beenden \vspace{1em} \noindent
The first part of Theorem \ref{t1} in particular states that these
matrix elements may be viewed as meromorphic functions of the
formal variables. Thus vertex algebras can be seen as a rigorous
formulation of meromorphic conformal field theories.
Note that the second part of the theorem should be interpreted as
``duality (crossing symmetry) of the 4-point correlation function on
the Riemann sphere''. It establishes a precise formulation of an
operator product expansion in two-dimensional conformal field theory
\cite{Bank88},\cite{Gins89},\cite{Godd89b} in the sense that
$\VP{1}\VF{2}$ agrees with $\sum_{n\inZ}(z_1-z_2)^{-n-1}\V(\p_n\f,z_2)$
as operator valued rational functions.The theorem then also ensures
that this operator product expansion involves only finitely many
singular (at ``$z_1=z_2$'') terms.
It is worth mentioning that Theorem \ref{t1} contains the full
information about the Jacobi identity, i.e. one can derive the latter
starting from the principles of locality and duality. Even more is true.
Using products of three vertex operators, duality follows from locality,
(\ref{31}), (\ref{32}) and the axioms for vertex algebras except for the
Jacobi identity. In particular, in the definition of a vertex algebra
the Jacobi identity may be replaced by the principle of locality,
(\ref{31}) and (\ref{32}). Proofs of these statements can be found in
\cite{FrHuLe93} where also the generalization of above notion of duality
to arbitrary $n$-point functions is presented.

If we regard our formal variables as {\it complex} variables then the
formal expansions of rational functions that we have been discussing
converge in suitable domains. The matrix elements in (\ref{46}) and
(\ref{47}) converge to a common rational function in the disjoint
domains $|z_1|>|z_2|>0$ and $|z_2|>|z_1|>0$, respectively. The matrix
elements in (\ref{48}) and (\ref{49}) for $z_0=z_1-z_2$ converge to a
common rational function in the domains $|z_2|>|z_1-z_2|>0$ and
$|z_1|>|z_2|>0$, respectively, and in the common domain
$|z_1|>|z_2|>|z_1-z_2|>0$ these two series converge to the common
function.

Finally, we would like to discuss the space $\F^\prime$ and the pairing
$\lx\_|\_\rx$ in more detail. We can assign to each state $\p\in\F$ a
{\bf contragredient vertex operator} $\V^*(\p,z)=\sum\p_n^*z^{-n-1}\in
(\End\F^\prime)[\![z,z^{-1}]\!]$ by the condition
\[ \lx\V^*(\p,z)\chi^*|\f\rx=\lx\chi^*|
      \V\left(\e^{z\Le}(-z^{-2})^{\Lo}\p,z^{-1}\right)\f\rx \]
It is crucial to observe that the formal sum on the right hand side
in general does not exist (check!) unless $(\Le)^n\p=0$ for $n$ large
enough which is assured if the spectrum of $\Lo$ is bounded below. For
the remainder of this subsection we will therefore assume that \vertex
is a vertex operator algebra.
Without giving the precise definition of a module for a vertex operator
algebra we just state that with this definition $(\F^\prime,\V^*)$
                                                              becomes a
\vertex-module (For a proof and the relevant definitions see \cite{
FrHuLe93}).If we furthermore have a grading--preserving linear isomor%
phism $F:\F\to\F^\prime$ then this amounts to choosing a nondegenerate
bilinear form $(\_,\_)$ on $\F$ as $(\chi,\f):=\lx F(\chi)|\f\rx$ with
                                                                     the
{\bf adjoint vertex operator} defined by
\beq \V^\dagger(\p,z)=\sum_{n\inZ}\p_n^\dagger z^{-n-1}:=
     \V\left(\e^{z\Le}(-z^{-2})^{\Lo}\p,z^{-1}\right)
     \in(\End\F)[\![z,z^{-1}]\!] \labelx{dual} \eeq
such that $(\Vp\chi,\f)=(\chi,\V^\dagger(\p,z)\f)$.\\
This definition of an adjoint vertex operator is quite close to the one
familiar to physicists as one immediately sees by calculating explicitly
the adjoint vertex operator associated with a quasi-primary state $\p$.
\baro \V^\dagger(\p,z)&=&(-1)^{\Delta_\p} z^{-2\Delta_\p}\V(\p,z^{-1})
      \qquad \mbox{since $\p$ quasi-primary}\\
          &=&(-1)^{\Delta_\p}\sum_{n\inZ}\p_nz^{n+1-2\Delta_\p} \earo
i.e.
\beq \p_n^\dagger=(-1)^{\Delta_\p}\p_{-n+2\Delta_\p-2}\qquad\forall
     n\in\Z \labelx{qpdual}\eeq
With the shifted grading $\p_{(n)}\equiv\p_{n+\Delta_\p-1}$ this reads
\[ \p_{(n)}^\dagger=(-1)^{\Delta_\p}\p_{(-n)}\qquad\forall n\in\Z \]
In particular, we observe that the vacuum vertex operator (identity) is
selfadjoint and that the Virasoro generators satisfy the well-known
relation $\Ln^\dagger=\Lna$ or, in terms of the ``stress energy tensor''
, $\V^\dagger(\w,z)=\frac1{z^4}\V(\w,z^{-1})$ \cite{Bank88}. Hence we
obtain for any two homogeneous elements $\chi\in\Fo{m},\ \f\in\Fo{n}$,
\[ (m-n)(\chi,\f)=(\Lo\chi,\f)-(\chi,\Lo\f)=0 \]
i.e. the homogeneous subspaces $\Fo{n},n\in\Z,$ are orthogonal to each
other with respect to this bilinear form,
\[ (\Fo{m},\Fo{n})=0\If{}m\neq n \]
For completeness we just mention (and encourage the reader to do the
straightforward calculation) that for an $\su(1,1)$-descendant state
$\p^{(N)}\equiv\frac1{N!}(\Lt)^N\p=\p_{-N-1}\1$ the adjoint is given by
\[ (\p_n^{(N)})^\dagger=\sum_{i=0}^N(-1)^{\Delta_\p+i}
   {2\Delta_\p+i \choose N-i}\p_{-n+N+i+2\Delta_\p-2}^{(i)} \]
in agreement with (\ref{qpdual}) for $N=0$.\\
To check whether adjointness satisfies the involution property $\V^{
\dagger\dagger}=\V$ we note that the commutation relation $[\Lo,\Ln]=
-n\Ln$ yields $z_0^{\Lo}\Ln z_0^{-\Lo}=z_0^{-n}\Ln$ which by iteration
gives us the conjugation formula
\[ z_0^{\Lo}\e^{z\Ln}z_0^{-\Lo}=\e^{z_0^{-n}z\Ln} \]
so that we have indeed
\[   \V\left(\e^{z^{-1}\Le}(-z^2)^{\Lo}\e^{z\Le}(-z^{-2})^{\Lo}\p,z
     \right)=\Vp\qquad\forall\p\in\F \]
It is by no means obvious from the definition that the bilinear form is
symmetric. To establish symmetry we first note that $(\chi,\f)=\Res{z}{
z^{-1}(\V(\chi,z)\1,\f)}$ by (\ref{36}). Therefore it is sufficient to
prove that $(\V(\chi,z)\1,\f)=(\f,\V(\chi,z)\1)$.Now,
\baro
(\V(\chi,z)\1,\f)
&=&(\1,\V\left(\e^{z\Le}(-z^{-2})^{\Lo}\chi,z^{-1}\right)\f)
                                        \by{definition}\\
&=&(\1,\e^{z^{-1}\Lt}\V(\f,-z^{-1})\e^{z\Le}(-z^{-2})^{\Lo}\chi)
                                        \by{skew-symmetry}\\
&=&(\V\left(\e^{-z^{-1}\Le}(-z^2)^{\Lo}\f,-z\right)\1,\e^{z\Le}
                    (-z^{-2})^{\Lo}\chi)  \by{involution}\\
&=&(\V(\1,z)\e^{-z^{-1}\Le}(-z^2)^{\Lo}\f,(-z^{-2})^{\Lo}\chi)
                                        \by{skew-symmetry}\\
&=&(\f,(-z^2)^{\Lo}\e^{-z^{-1}\Lt}(-z^{-2})^{\Lo}\chi)
                                        \by{definition}\\
&=&(\f,\V(\chi,z)\1) \by{conjugation} \earo

\subsection{Algebras of primary fields of weight one}
We shall provide a certain subspace of the Fock space $\F$ with the
structure of a Lie algebra.(cf. \cite{Borc86},\cite{Borc92},\cite{FLM88}
) \\ Looking at the skew-symmetry property (\ref{35}) we can define a
product by \beq [\p,\f]:=\p_0\f \labelx{50} \eeq
which is antisymmetric on the subspace \Flie . Then, as already
mentioned at the end of \ref{3.1}, the classical Jacobi identity for
Lie algebras,
\beq [[\p,\f],\xi]+[[\f,\xi],\p]+[[\xi,\p],\f]=0 \labelx{50b} \eeq
follows from the Jacobi identity for vertex algebras. \\
Another glimpse at skew-symmetry shows that the Lie algebra
\Flie is also equipped with a symmetric product by
\beq \lx\p,\f\rx:=\p_1\f \labelx{51} \eeq
To investigate possible \Flie-invariance of $\lx\_,\_\rx$ we
note that
\baro \lx[\p,\f],\xi\rx &\equiv& (\p_0\f)_1\xi \\
      &=& \p_0(\f_1\xi)-\f_1(\p_0\xi)\by{(\ref{20})} \\
      &\equiv& [\p,\lx\f,\xi\rx]-\lx\f,[\p,\xi]\rx        \earo
Hence the product $\lx\_,\_\rx$ is in general {\it not} \Flie         -
invariant unless we make further assumptions.

For that purpose let us restrict our attention to the piece of conformal
weight one, i.e. to $\Fo{1}$. Then the space
\[ \Fo{1}\Big/(\Lt\F\cap\Fo{1})=
                               \Fo{1}\big/\Lt\Fo{0} \]
is a subalgebra of the Lie algebra \Flie and, by (\ref{33}),
\[ \lx\f,\xi\rx\in\Fo{0}\for{all }\f,\xi\in\Fo{1} \]
If we now assume that $\Fo{0}$ is one-dimensional then $\lx\f,\xi\rx$
, $\f,\xi\in\Fo{1}$, is a scalar multiple of the vacuum and
$\Lt\Fo{0}=0$ by (\ref{28}). Thus $[\p,\lx\f,\xi\rx]$ is proportional
to $\p_0\1$ which vanishes because of (\ref{37}) and we have indeed
established invariance of the scalar product.\\
In some cases we can say even more. Let us look again at the commutator
formula (\ref{20c}):
\baro [\p_m,\f_n] &=&
                      \sum_{i\ge0}{m \choose i}(\p_i\f)_{m+n-i} \\
      &=& (\p_0\f)_{m+n}+m(\p_1\f)_{m+n-1}+
                      \sum_{i\ge2}{m \choose i}(\p_i\f)_{m+n-i} \\
      &=& ([\p,\f])_{m+n}+m\lx\p,\f\rx\1_{m+n-1}+
                      \sum_{i\ge2}{m \choose i}(\p_i\f)_{m+n-i} \\
      &=& ([\p,\f])_{m+n}+m\lx\p,\f\rx\delta_{m+n,0}{\rm id}_\F+
 \sum_{i\ge2}{m \choose i}(\p_i\f)_{m+n-i} \by{(\ref{17})} \earo
for $\p,\f\in\Fo{1}$. Since all states $\p_i\f$, $i\ge2$, which occur
in the sum on the right-hand side have negative conformal weight we
conclude:
\begin{theo} {\bf:} \labelx{t2} \\
If the weight zero piece , $\Fo{0}$, of a vertex algebra \vertex
is one-dimensional and the spectrum of the operator $\Lo$ is nonegative
then the weight one piece, $\Fo{1}$, is a Lie algebra with
antisymmetric product $[\p,\f]:=\p_0\f$ and $\Fo{1}$-invariant bilinear
form $\lx\p,\f\rx:=\p_1\f$. The space
\[ \hat{\Fo{1}}:=\{\p_n|\p\in\Fo{1},n\in\Z\}\oplus\{{\rm id}_\F\} \]
provides a representation of the affinization of $\Fo{1}$ on $\F$ by
\[ [\p_m,\f_n]=([\p,\f])_{m+n}+m\lx\p,\f\rx\delta_{m+n,0}{\rm id}_\F \]
In particular, $\Fo{1}$ may be identified with the Lie algebra of
operators $\{\p_0|\p\in\Fo{1}\}$ on $\F$ since in the adjoint
representation, $\Fo{1}$ acts on itself as operators
$\p_0$. \end{theo} \vspace{1em}

It is quite interesting that in physical applications such as string
theory, two-dimensional statistical systems and two-dimensional quantum
field theories {\it physical} considerations lead to the same condition
on the spectrum of $\Lo$ as in Theorem \ref{t2}. In such theories $\Lo$
is identified with the Hamiltonian so that above condition immediately
translates into the postulate of the positivity of the energy.(see
\cite{GodOli86}, e.g.)

The Lie algebra $\Fo{1}\big/\Lt\Fo{0}$ is still too large for further
investigations. Equation (\ref{41}) tells us that if $\p$ is a primary
state of weight one then the corresponding operator $\p_0$ commutes
with the Virasoro algebra, so it preserves all the physical subspaces
$\Ph{n}$. In other words, it maps physical states into physical states.
Hence it is natural to look in detail at the Lie algebra
\[ \g:=\Ph{1}\Big/(\Lt\Fo{0}\cap\Ph{1}) \]
Note that if $\p=\Lt\f\in\Lt\Fo{0}$ for some $\f\in\Fo{0}$ then
$\p_0=(\f_{-2}\1)_0=0$ by (\ref{37}), (\ref{20b})(with $l=-2,\ n=0$)
and (\ref{17}).\\
Again, if $\Fo{0}$ is one-dimensional then
                                          $\Ph{1}$ is a Lie algebra with
antisymmetric product and invariant bilinear form defined as above. \\
On the other hand let us consider the case where $\Lo$ has nonegative
spectrum. Then
\[ \Lt\Fo{0}\subset\Ph{1} \]
because
\[ \Ln\Lt\p\in\Fo{1-n},\quad\Le\Lt\p=\Lt\Le\p\in\Lt\Fo{-1}\for{}\p\in
   \Fo{0} \]
and we find that $\Ph{1}\big/\Lt\Fo{0}$ is a Lie algebra.

Borcherds \cite{Borc86},\cite{Borc88},\cite{Borc92} was led to his
definition of generalized Kac-Moody algebras precisely by Lie algebras
of type $\Ph{1}\big/(\Lt\Fo{0}\cap\Ph{1})$ for vertex algebras
associated with even Lorentzian lattices.

When defining the Lie algebra \Flie we had to divide out the
space $\Lt\F$ for mathematical reasons. Surprisingly, there is also a
physical explanation for that procedure.(cf. \cite{GSW88}) Suppose that
$\F$ is equipped with an inner product (\_,\_) such that the operator
$\Lna$ is the adjoint of $\Ln$ (cf. end of last subsection!). Then
\[ (\Lt\f,\p)=(\f,\Le\p)=0\for{all }\f\in\F,\p\in\Ph{n},n\in\Z \]
i.e. the space $\Lt\F$ is orthogonal to all physical states. In
particular, $\Lt\Fo{0}\cap\Ph{1}$ consists of "null" physical states,
physical states orthogonal to all physical states including themselves.
Hence the Lie algebra $\g$ is obtained from $\Ph{1}$ by dividing out
null physical states.\\
It turns out that there are additional null physical states in $\Ph{1}$
if and only if the central charge takes the critical value $c=26$,
namely the space $(\Lza+\frac32\Lt^2)\Ph{-1}$. Evidently, this space is
annihilated by $\Ln$ for $n\ge3$. Furthermore
\baro \Le(\Lza+{\textstyle\frac32}\Lt^2)\Ph{-1} &=&
      [\Le,\Lza+{\textstyle\frac32}\Lt^2]\Ph{-1}\\
      &=& 3(\Lt+\Lo\Lt+\Lt\Lo)\Ph{-1} \\ &=& 0 \\[.5em]
      \Lz(\Lza+{\textstyle\frac32}\Lt^2)\Ph{-1} &=&
      [\Lz,\Lza+{\textstyle\frac32}\Lt^2]\Ph{-1}\\
      &=& \Big(4\Lo+{\textstyle\frac{c}2}+{\textstyle\frac92}
                                     (\Lo\Lt+\Lt\Lo)\Big)\Ph{-1} \\
      &=& 0\If{and only if }c=26 \earo
The existence of these additional null physical states is used in the
proof of the No-ghost-theorem.\cite{GodTho72},\cite{GSW88}

In applied conformal field theory one usually encounters the situation
that the spectrum of $\Lo$ is bounded below and the Fock space $\F$ is
equipped with a positive definite inner product $(\_,\_)$. Then a
standard argument (see e.g. \cite{Gins89}) shows that in this case
$\F$ splits up into a direct sum of $\su(1,1)$ highest-weight
representations generated by some basis of the quasi-primary states.
Since for any quasi-primary state $\p\in\F$ the $\su(1,1)$
representation space $[\p]\subset\F$ generated from $\p$ is spanned by
the elements of $\{(\Lt)^N\p|N\in\N\}$ we may identify \Flie
with the set of quasi-primary states in the Fock space $\F$.

\subsection{Cross-bracket and the algebra of fields of weight two}
\labelx{3.5}
Before we leave the axiomatics of vertex algebras and turn to examples
let us investigate the case where the vertex algebra \vertex contains
{\it no} states of weight one.(see \cite{FLM88}) \\
Define a product by \[ \p\times\f:=\p_1\f\for{}\p,\f\in\Fo{2} \]
which is symmetric but non-associative on the piece $\Fo{2}$ in view of
(\ref{35}) with $\Lt\Fo{1}=0$. Note that $\frac12\w$ provides an
identity element on $\Fo{2}$,
\[ {\textstyle \frac12}\w\times\p={\textstyle \frac12}\w_1\p=
{\textstyle \frac12}\Lo\p=\p\qquad\forall\p\in\Fo{2} \]

If we assume that $\Fo{0}$ is one-dimensional then $\p_3\f$ for
$\p,\f\in\Fo{2}$ is a scalar multiple of the vacuum by (\ref{33}) so
that \[ \lx\p,\f\rx:=\p_3\f \] gives us a symmetric bilinear form on
$\Fo{2}$. Moreover, this form is associative in the sense that
\[ \lx\f,\p\times\xi\rx=\lx\f\times\p,\xi\rx\for{}\p,\f,\xi\in\Fo{2} \]
This can be seen most easily by setting $l=m=n=1$ in (\ref{20}):
\[ \p_2(\f_2\xi)-\f_2(\p_2\xi)-\p_1(\f_3\xi)+\f_3(\p_1\xi)=(\p_1\f)_3\xi
   +(\p_2\f)_2\xi \]
Since $\Fo{1}=0$ the first two terms on the left-hand side and the last
term on the right-hand side vanish while $\p_1(\f_3\xi)$ is proportional
to $\p_1\1$ which is zero because of (\ref{37}).

We define the {\bf cross-bracket} as follows:
\[ [\p_m\times_1\f_n]\equiv
   [\p\times_1\f]_{mn}
   :=[\p_{m+1},\f_n]-[\p_m,\f_{n+1}]\for{}\p,\f\in\Fo{2} \]
This looks a kind of awkward but turns out to be quite interesting as
soon as one reminds the Jacobi identity in components, (\ref{20}),
which gives for $l=1$, $\p,\f\in\Fo{2}$:
\baro
   [\p\times_1\f]_{mn} &=&
    \sum_{i\ge0}{m \choose i}(\p_{i+1}\f)_{m+n-i} \\
  &=& (\p_1\f)_{m+n}+m(\underbrace{\p_2\f}_{=0})_{m+n-1}+\frac12m(m-1)
      (\p_3\f)_{m+n-2}+\sum_{i\ge3}{m \choose i}(\p_{i+1}\f)_{m+n-i} \\
 &=&(\p\times\f)_{m+n}+\frac12m(m-1)\lx\p,\f\rx\delta_{m+n,1}{\rm id}_\F
        +\sum_{i\ge3}{m \choose i}(\p_{i+1}\f)_{m+n-i} \\     \earo
Since the sum on the right-hand side involves only negative weight terms
we have arrived at the following result:
\begin{theo} {\bf:} \labelx{t3} \\
Let \vertex be a vertex algebra. If the weight zero piece is
one-dimensional, the weight one piece is empty and the spectrum of the
operator $\Lo$ is nonegative then the weight two piece, $\Fo{2}$, is a
non-associative algebra with symmetric product $\p\times\f:=\p_1\f$ and
associative bilinear form $\lx\p,\f\rx:=\p_3\f$. The space
\[ \hat{\Fo{2}}:=\{\p_n|\p\in\Fo{2},n\in\Z\}\oplus\{{\rm id}_\F\} \]
provides a representation of the commutative affinization of $\Fo{2}$
on $\F$ by the cross-bracket,
\[ [\p_m\times_1\f_n]=(\p\times\f)_{m+n}+\frac12m(m-1)
 \lx\p,\f\rx\delta_{m+n,1}{\rm id}_\F \] \end{theo} \vspace{1em}
Of course, this cross-bracket is quite a nice algebraic structure in
our vertex algebra but immediately the question arises whether such a
commutative non-associative algebra really exists. In fact, the so
called Moonshine Module (see also Subsection \ref{5.3}) constructed by
Frenkel, Lepowsky, and Meurman {\it is} a vertex operator algebra that
satisfies all the assumptions of Theorem \ref{t3} and it turns out that
then $\Fo{2}$ is precisely the 196884--dimensional {\bf Griess algebra}
which possesses the Monster group, F$_1$, as its full automorphism
group.\cite{Grie82},\cite{FrLeMe85},\cite{FrLeMe86},\cite{Borc86}
\subsection{Symmetry products}
The commutator and the cross-bracket of the last two subsections can be
embedded in an infinite family of symmetry products by using the Jacobi
identity in the following way \cite{FLM88}:
Take $\Res{z_0}{z_0^n(Jacobi\ identity)}$ and define for $l\in\Z$
\barr [\VP1\times_l\VF2]&:=&\Res{z_0}{z_0^l\D210\VV{z_0}} \nonumber \\
   &=&(z_1-z_2)^l\VP1\VF2-(-z_2+z_1)^l\VF2\VP1 \labelx{sp1} \earr
which is expressed in modes as
\barr [\p\times_l\f]_{mn}&:=&\sum_{i\ge0}(-1)^i{l \choose i}\left(
      \p_{m+l-i}\f_{n+i}-(-1)^l\f_{n+l-i}\p_{m+i}\right) \nonumber \\
     &=&\sum_{i\ge0}{m \choose i}(\p_{l+i}\f)_{m+n-i} \labelx{sp2} \earr
It is clear that these products are symmetric for odd $l$ and
alternating for even $l$. The symmetric product $\times_1$ is precisely
the cross-bracket used to construct the affinization of the Griess
algebra in Theorem \ref{t3} while the alternating product $\times_0$
yields nothing but the commutator for the modes in Theorem \ref{t2},
$[\p\times_0\f]_{mn}=[\p_m,\f_n]$.
As the interpretation of the other symmetry
products regards so far we can only associate with the product
$\times_{-1}$ a well-known feature of conformal field theory. Recall
that the commutator of two fields is completely determined by the
singular part of the operator product expansion. The regular part of
the latter encodes the normal ordered product of two fields. Thus the
product $\times_{-1}$ which differs from $\times_0$ by a factor $(z_1-
z_2)^{-1}$ (more precisely, by a factor $z_0^{-1}$ in $\Res{z_0}{\ldots}
$) might be a good guess for defining normal ordered products in a
vertex algebra. For this purpose one would usually consider the
(algebraic) limit $z_1\to z_2$ of $[\VP1\times_{-1}\VF2]$ which
unfortunately does not exist. Hence we start with the following
definition of {\bf normal ordered product}:
\barr \Ord\Vp\Vf\Ord&:=&
      \sum_{n\inZ}[\p\times_{-1}\f]_{0n}z^{-n-1} \nonumber \\
      &=&\sum_{n\inZ}(\p_{-1}\f)_nz^{-n-1} \nonumber \\
      &=&\V(\p_{-1}\f,z) \labelx{sp3} \earr
In the following we will also refer to the product $\p_{-1}\f$ as
normal ordered product of states.
At first sight this does not look like the standard normal ordered
product of fields in conformal field theory. However, using the mode
expansion (\ref{sp2}) we can rewrite the definition as
\barr \Ord\Vp\Vf\Ord
      &=&\sum_{n\inZ}\sum_{i\ge0}(\p_{-i-1}\f_{n+i}+
                        \f_{n-i-1}\p_i)z^{-n-1} \nonumber \\
      &=&\sum_{n\inZ}\sum_{i\ge0}\ord\p_{-i-1}\f_{n+i}
                        \ord z^{-n-1} \labelx{sp4} \earr
where we introduced the normal ordering of the modes,
\[ \ord\p_{-i-1}\f_{n+i}\ord:=\cases{\p_{-i-1}\f_{n+i} & if $i\ge0$ \cr
                                 \f_{n+i}\p_{-i-1} & if $i<0$ } \]
And this is indeed the familiar normal ordered product of modes \cite{
Godd89b} if we employ the standard shifted grading $\p_{(n)}\equiv\p_
{n+\Delta_\p-1}$.\\
In general one expects the normal ordered product of bosonic fields to
be commutative. Since skew-symmetry (\ref{35}) forces
$\p_{-1}\f=\f_{-1}\p$ on \Flie, we therefore should restrict
above definition of normal ordered product to the quotient space
\Flie.This has the nice effect of automatically projecting
the normal ordered products of quasi-primary fields onto the space of
quasi-primary fields. The idea is to subtract expressions of the form
$(\Lt)^i(\p_{i-1}\f)$, $i\ge1$, from $\p_{-1}\f$ such that one ends
up with a quasi-primary state. Indeed, we found that the projected
normal ordered product of two quasi-primary states $\p,\f$ is given by
\[ [\p*_{-1}\f]:=\p_{-1}\f+\sum_{i\ge1}\frac{(-1)^i}{i!}
       {2\Delta_\p-1 \choose i}{2(\Delta_\p+\Delta_\f-1) \choose i}^{-1}
          (\Lt)^i(\p_{i-1}\f) \]
i.e. $[\p*_{-1}\f]$ is a quasi-primary state of weight $\Delta_\p+\Delta
_\f$. This formula is similar to those given in \cite{Bowc91} and
\cite{Nahmetal91}. For the sake of completeness we would like to
mention that this projection onto quasi-primary states generalizes to
any product $\p_n\f$, $n\in\Z$, of two quasi-primary states, i.e.
\[ [\p*_l\f]:=\sum_{i\ge0}\frac{(-1)^i}{i!}
              p_i^{(l)}(\Delta_\p,\Delta_\f)(\Lt)^i(\p_{i+l}\f) \]
with
\[ p_i^{(l)}(\Delta_\p,\Delta_\f):=
{2\Delta_\p-2-l \choose i}{2(\Delta_\p+\Delta_\f-2-l) \choose i}^{-1} \]
is a quasi-primary state of weight $\Delta_\p+\Delta_\f-l-1$. If we
calculate the corresponding vertex operator using the translation
axiom (\ref{tr}) we find
\[ \V([\p*_l\f],z)=\sum_{m\inZ}\sum_{i\ge0}
       p_i^{(l)}(\Delta_\p,\Delta_\f){m \choose i}(\p_{l+i}\f)_{m-i} \]
Surprisingly, we observe that these projected products are just the
``old'' symmetry products (\ref{sp2}) for $n=0$ where each term in the
sum is weighted with an additional polynomial factor
$p_i^{(l)}(\Delta_\p,\Delta_\f)$. Moreover we can prove that the
projected products also inherit the symmetry properties from the
original ones, i.e. $[\p*_l\f]=(-1)^{l+1}[\f*_l\p]$.

It is quite interesting that the normal ordered product (\ref{sp3})
turns out to be also associative if we consider the quotient space
\Fger
 To see this we first note that $\Lt\F= \Fo{-2}\1\subset \Fo{-2}\F$
by (\ref{37}). Additionally we have for $n\le-2$,
\baro \p_n\f&=&\p_n(\f_{-1}\1)=[\p_n,\f_{-1}]\1+\f_{-1}(\p_n\1) \\
      &=&\sum_{i\ge0}{n \choose i}(\p_i\f)_{n-1-i}\1+
         \frac1{(-n-1)!}\f_{-1}((\Lt)^{-n-1}\p) \\
      &=&\underbrace{\sum_{i\ge0}\frac1{(i-n)!}{n \choose i}
         \left((\Lt)^{i-n-1}(\p_i\f)\right)_{-2}\1}_{\in\Fo{-2}\1}+
         \underbrace{\frac1{(-n-1)!}\f_{-1}
         \left(((\Lt)^{-n-2}\p)_{-2}\1\right)}_{\in\Fo{-2}\F} \earo
where the last term lies in $\Fo{-2}\F$ because of
\[ \f_{-1}(\xi_{-2}\1)=\xi_{-2}\f+\sum_{i\ge0}\frac{(-1)^i}{(i+2)!}
   \left((\Lt)^{i+1}(\f_i\xi)\right)_{-2}\1\qquad\forall\f,\xi\in\F \]
This gives us associativity of the normal ordered product,
\baro (\p_{-1}\f)_{-1}\chi-\p_{-1}(\f_{-1}\chi)&=&
              \sum_{i\ge0}\left\{\p_{-1-i}(\f_{-1+i}\chi)
              +\f_{-2-i}(\p_i\chi)\right\}-\p_{-1}(\f_{-1}\chi) \\
              &=&0 \qquad\mbox{on \Fger} \earo
The fact that $\p_n\f\in\Fo{-2}\F$ for $n\le-2$ in particular implies
$\Ln\F\subset \Fo{-2}\F$ for $n\le-3$ which allows us to give a nice
interpretation of the quotient space \Fger. Consider a
conformal family $[\p]$ associated with a primary state $\p\in \F$. It
is spanned by elements of the form \cite{BePoZa84}
\[ (\Lt)^{i_1}(\Lza)^{i_2}\ldots(\Lna)^{i_n}\p\qquad n\ge1,\ i_1,\ldots
   ,i_n\ge0 \]
Hence only the subspace spanned by the states $(\Lza)^N\p$, $N\in\N$,
survives when the conformal family is projected onto \Fger.
Moreover, $\Lza\equiv\w_{-1}$ so that these states are just multiple
normal ordered products of the Virasoro vector with the conformal
highest weight vector $\p$.\\
We conclude:
The quotient
space \Fger with the induced normal ordered product
$\p_{-1}\f$ carries the structure of a commutative associative algebra.
If the Fock space $\F$ splits up into a direct sum of highest weight
representations of the Virasoro algebra then \Fger can be
identified with the span of primary states, Virasoro vector, and
multiple normal ordered products of the latter with the primary
states.

Above quotient space plays a special role when vertex algebras on the
torus are considered i.e. when the notion of modular invariance is
built into the framework of vertex algebras \cite{Zhu90}.
A vertex operator algebra \vertex is said to satisfy the
{\bf finiteness condition} if the quotient space \Fger
is finite-dimensional. A vertex operator algebra is called
rational if it has only finitely many irreducible representations and
every finitely generated representation is completely reducible.
In fact, Zhu \cite{Zhu90} proved that if a rational vertex operator
algebra satisfies the finiteness condition then the linear span of the
characters ${\rm tr}\,q^{\Lo-\frac c{24}}$ of its irreducible
representations is modular invariant with respect to SL(2,$\Z$). It is
conjectured that rational vertex operator algebras automatically
satisfy the finiteness condition.

Finally we would like to mention that above discussion of the quotient
space \Flie is especially useful for explaining how Gerstenhaber
algebras arise in the context of super vertex algebras. To see
this, one supposes that an odd operator $Q$ satisfying $Q^2=0$, acts
on the vertex algebra \vertex, and that $Q$ can be represented as the
zero mode of a vertex operator $\V(\sig,z)$ associated with an odd
``ghost'' state $\sig$ of weight minus one, i.e. $Q=\sig_0$.
Furthermore one assumes that there is an odd ``antighost'' state
$\bet$ of weight two such that
$\Ln\equiv\w_{n+1}= (Q\bet)_{n+1}\equiv (\sig_0\bet)_{n+1}\ \forall n$.
It is proved in \cite{LiaZuc92} and \cite{PenSch92} that the superspace
$\ker Q/{\rm im}\,Q$, the cohomology of $Q$, can be equipped with the
structure of a Gerstenhaber algebra. It turns out that the dot product
in this superspace is precisely given by $\p_{-1}\f$. We observe that
the structure of \Fger is closely related to that of the cohomology
of the nilpotent operator $Q$ if we collect above formulas for
the product $\p_{-1}\f$ and add the properties of the bracket
operation (\ref{50}),(\ref{50b}),(\ref{20d}),(\ref{33}) to obtain
\begin{theo} {\bf:}\\
The dot product $\p\cdot\f:= \p_{-1}\f$ and the bracket $[\p,\f]:=
\p_0\f$ enjoy the following properties on the quotient space \Fger,
\beginnen
  \item[(i)] $\p\cdot\f=\f\cdot\p$
  \item[(ii)] $(\p\cdot\f)\cdot\xi=\p\cdot(\f\cdot\xi)$
  \item[(iii)] $[\p,\f]=-[\f,\p]$
  \item[(iv)] $[[\p,\f],\xi]+[[\f,\xi],\p]+[[\xi,\p],\f]=0$
  \item[(v)] $[\p,\f\cdot\xi]=[\p,\f]\cdot\xi+\f\cdot[\p,\xi]$
  \item[(vi)] $[\_,\_]\ :\ \Fgerq{m}\times\Fgerq{n}\longrightarrow
                          \Fgerq{m+n-1}$
\beenden
\end{theo}
Thus the quotient space \Fger may be thought of as part of some
Gerstenhaber algebra \cite{LiaZuc92}. In fact, the ghost number one
part of the cohomology of $Q$ corresponds precisely to the primary
states of weight one which are also contained in \Fger. We do not
know yet how significant the resemblance of these two algebraic
structures is.

\section{Vertex algebras associated with even lattices}
\subsection{Statement of the theorem}
It is by no means obvious that nontrivial examples of vertex (operator)
algebras exist. However, a class of vertex algebras is provided by the
following result.(see \cite{Borc86},\cite{FLM88},\cite{Dong1})
\begin{theo}\labelx{t4} {\bf:}\\
Associated with each nondegenerate even lattice $\L$ there is a vertex
algebra \vertex. If in addition $\L$ is positive definite then
\vertex has the structure of a vertex operator algebra.
\end{theo} \vspace{1em}
In fact, above examples of vertex algebras gave rise to the notion and
the abstract definition of vertex algebras. Frenkel and Zhu \cite
{FreZhu92} constructed vertex operator algebras corresponding to the
highest weight represenations of affine Lie algebras with the highest
weights being multiples of the highest weight of the basic representat%
ion. Up to now these two classes are essentially all known examples
of (untwisted) vertex operator algebras. Hence it is desirable
to find other classes of examples of vertex algebras to which the
general formalism may be applied.
At present, however, research is concentrating on representation theory
of vertex algebras \cite{FreZhu92},\cite{Dong1},\cite{Dong2},\cite{Don%
g3},\cite{DonMas1},\cite{DonMas2}, generalizations of vertex algebras
\cite{DonLep1} and the geometric interpretation of vertex algebras
\cite{Huan90},\cite{Huan91},\cite{HuaLep93}.

The rest of this section we will be concerned with the explicit
construction of the vertex algebra stated above.
\subsection{Fock space and vertex operators} \labelx{4.1}
For physical motivations of the construction below the reader may
consult the beautiful articles \cite{GodOli85},\cite{Godd86},\cite{God%
Oli86} or the comprehensive review \cite{LeScheWa89}.\\
Let $\L$ be an even lattice of rank $d<\infty$ with a symmetric
nondegenerate $\Z$-valued $\Z$-bilinear form $\_\cdot\_$ and
corresponding metric tensor $\eta^{\mu\nu}$, $1\le\mu,\nu\le d$ ($\L$
even means that $\r^2\in2\Z$ for all $\r\in\L$). The vertex algebra
\vertex which we shall construct can be thought of as a bosonic
string theory with $d$ spacetime dimensions compactified on a torus.
Thus $\L$ represents the allowed momentum vectors of the theory\footnote
{At this point we are rather sloppy since we should work with the
 complexified lattice
 $\L_{\mbox{\bbsss C}}\equiv\L\otimes_{\mbox{\bbsss Z}}\mbox{\bbss C}$
 rather than with the real lattice $\L$ itself. However, we believe
 that this subtlety is not essential for a physicist's understanding of
 the general construction.}.\\
Introduce orthonormal vectors ("zero mode states") $\Pr$, $\r\in\L$,
\[ (\Pr,\Ps)=\delta_{\r\s} \]
and oscillators $\alm$, $m\in\Z$, $1\le\mu\le d$, satisfying the
commutation relations
\[ [\alm,\aln]=m\eta^{\mu\nu}\delta_{m+n,0}, \]
the hermiticity conditions
\[ (\alm)^\dagger=\alpha^\mu_{-m}, \]
and acting on zero mode states by
\baro \alm\Pr&=&0\If{}m>0\\ p^\mu\Pr&=&r^\mu\Pr \earo
where $p^\mu\equiv\al^\mu_0$ and $r^\mu$ are the components of
$\r\in\L$. While the operators $\alm$ for $m>0$ by definition act as
annihilation operators, the creation operators $\alm$, $m<0$, generate
the Fock space from the zero mode states. For convenience let us define
\[ \r(m):=\sum_{\mu=1}^dr_\mu\alm\equiv\r\cdot\alf_m \]
for $\r\in\L$, $m\in\Z$, such that
\beq [\r(m),\s(n)]=m(\r\cdot\s)\delta_{m+n,0} \labelx{60} \eeq
We denote the $d$-fold Heisenberg algebra spanned by the oscillators
by \[ \hat{\bf h}:=\{\r(m)|\r\in\L,m\in\Z\} \]
and for the vector space of finite products of creation operators
($\equiv$ algebra of polynomials on the oscillators) we write
\[ S\left(\hat{\bf h}^-\right):=
   \bigoplus_{N\inN}\left\{\prod_{i=1}^N\r_i(-m_i)|\r_i\in\L,\ m_i>0\
                          {\rm for}\ 1\le i\le N\right\} \]
where "$S$" stands for "symmetric" because of the fact that the creation
operators commute with each other.\\
If we introduce formally position operators $q^\mu$, $\1\le\mu\le d$,
commuting with $\alm$ for $m\neq0$ and satisfying
\[ [q^\nu,p^\mu]=i\eta^{\mu\nu} \]
then we find
\[ \e^{i\r\cdot\qq}\Ps=\Psi_{\r+\s} \]
i.e. the zero mode states can be generated from the vacuum $\Psi_0$:
\[ \Pr=\e^{i\r\cdot\qq}\Psi_0 \]
Thus the operators $\e^{i\r\cdot\qq}$, $\r\in\L$, may be identified with
the zero mode states and form an abelian group which is called the group
algebra of the lattice $\L$ and is denoted by $\C[\L]$. One might expect
the full Fock space $\F$ of the vertex algebra to be
$S\left(\hat{\bf h}^-\right)\otimes\C[\L]$. However,
it turns out that we shall need to replace the group algebra $\C[\L]$
by something more delicate in order to adjust some signs (in the Jacobi
identity). We will multiply $\e^{i\r\cdot\qq}$ by a so called cocycle
factor $c_\r$ which is a function of momentum $\pp$. This means that it
commutes with all oscillators $\alm$ and satisfies the eigenvalue
equations \[ c_\r\Ps=\epsilon(\r,\s)\Ps \]
Furthermore we define operators $\e^\r:=\e^{i\r\cdot\qq}c_\r$ and impose
the conditions
\barr \e^\r\e^\s&=&\epsilon(\r,\s)\e^{\r+\s} \labelx{61} \\
      \e^\r\e^\s&=&(-1)^{\r\cdot\s}\e^\s\e^\r \labelx{62} \\
      \e^\r\e^{-\r}&=&1 \labelx{62aa} \\
      \e^{\bf0}&=&1 \labelx{62a} \earr
which is equivalent to requiring, respectively,
\baro \epsilon(\r,\s)\epsilon(\r+\s,\t)&=&
      \epsilon(\r,\s+\t)\epsilon(\s,\t)    \\
      \epsilon(\r,\s)&=&(-1)^{\r\cdot\s}\epsilon(\s,\r) \\
      \epsilon(\r,-\r)&=&1 \\
      \epsilon({\bf0},{\bf0})&=&1 \earo
It is not difficult to show that it is always possible to construct
cocycles with these properties. Note that every 2-cocycle $\epsilon:\L
\times\L\to\{\pm1\}$ corresponds to a central extension $\hat{\L}$ of
$\L$ by $\{\pm1\}$: \[ 1\to\{\pm1\}\to\hat{\L}\to\L\to1 \]
where we put $\hat{\L}=\{\pm1\}\times\L$ as a set and define a
multiplication in $\hat{\L}$ by
\[ (\rho,\r)*(\sigma,\s):=(\epsilon(\r,\s)\rho\sigma,\r+\s)
   \for{}\rho,\sigma\in\{\pm1\},\ \r,\s\in\L \]
We will take the twisted group algebra $\C\{\L\}$ consisting of the
operators $\e^\r$, $\r\in\L$, instead of $\C[\L]$. This means nothing
but working with the double cover $\hat{\L}$ of the lattice $\L$.\\
We summarize: The Fock space associated with the lattice $\L$ is defined
to be \[ \F:=S\left(\hat{\bf h}^-\right)\otimes\C\{\L\} \]
Note that the oscillators $\r(m)$, $m\ne0$, act only on the first tensor
factor, namely, creation operators as multiplication operators and
annihilation operators via the adjoint representation i.e. by
(\ref{60}). The zero mode operators $\al^\mu_0$, however, are only
sensible for the twisted group algebra,
\beq \r(0)\e^\s\equiv(\r\cdot\pp)\e^\s=(\r\cdot\s)\e^\s
     \for{all }\r,\s\in\L  \labelx{63} \eeq
while the action of $\e^\r$ on $\C\{\L\}$ is given by (\ref{61}).

We shall define next the (untwisted) vertex operators $\Vp$ for $\p\in
\F$. For $\r\in\L$ we introduce the formal sum
\beq \r(z):=\sum_{m\inZ}\r(m)z^{-m-1} \labelx{63a} \eeq
which is an element of $\hat{\bf h}[\![z,z^{-1}]\!]$ and may be regarded
as a generating function for the operators $\r(m)$, $m\in\Z$, or as a
``currents'' in contrast to the "states" in $\F$. It is convenient to
split the current $\r(z)$ into three parts:
\[ \r(z)=\r_<(z)+\r(0)+\r_>(z) \] where
\[ \r_<(z):=\sum_{m>0}\r(-m)z^{m-1},\qquad
   \r_>(z):=\sum_{m>0}\r(m)z^{-m-1} \]
We will employ the usual normal ordering procedure, i.e. colons indicate
that in the enclosed expressions, $q^\nu$ is written to the left of
$p^\mu$, as well as the creation operators are to be placed to the left
of the annihilation operators.\\
For $\e^\r\in\C\{\L\}$, we set
\beq \V(\e^\r,z):=\e^{\int\r_<(z)dz}\e^\r z^{\r(0)}
     \e^{\int\r_>(z)dz} \labelx{64} \eeq
using an obvious formal integration notation, i.e.
\baro \int\r_<(z)dz&:=&\sum_{m>0}\frac1m\r(-m)z^m \\
      \int\r_>(z)dz&:=&-\sum_{m>0}\frac1m\r(m)z^{-m} \earo
This can be written in a way more familiar to physicists by introducing
formally the Fubini-Veneziano field,
\[ Q^\mu(z)\equiv q^\mu-ip^\mu\ln z+i\sum_{m\inZ}\frac1m\alm z^{-m} \]
which really only has a meaning when exponentiated. We find
\[ \V(\e^\r,z)={\bf:}\e^{i\r\cdot{\bf Q}(z)}{\bf:}c_\r \]
Let $\displaystyle\p=\bigg(\prod_{j=1}^N\s_j(-n_j)\bigg)\otimes\e^\r$ be
a typical homogeneous element of $\F$ and define
\barr \Vp&:=&\ord\;\V(\e^\r,z)\prod_{j=1}^N\frac1{(n_j-1)!}
       \left(\dz\right)^{n_j-1}\s_j(z)\;\ord \labelx{65}\\
    &\equiv&i\;\ord\;\e^{i\r\cdot{\bf Q}(z)}\prod_{j=1}^N
         \frac1{(n_j-1)!)}\left(\dz\right)^{n_j}(\s_j\cdot{\bf Q}(z))
         \;\ord\;c_\r  \nonumber\earr
where we used $\dz(i\s\cdot{\bf Q}(z))=\s(z)$. \\
Extending this definition by linearity we finally obtain a
well-defined map
\baro \V:\F&\to&(\End\F)[\![z,z^{-1}]\!] \\
      \p&\mapsto&\sum_{n\inZ}\p_nz^{-n-1} \earo
\subsection{Regularity, vacuum, injectivity, conformal vector}
We shall prove the first four axioms in the definition of a vertex
algebra. \beginnen
\item \BF{Regularity}\\
Note that $\F$ contains only states with a finite number of creation
operators and the vertex operators are normal ordered expressions.
Having this in mind it is clear that $\p_n\f=0$ for $n$ large enough
(depending on $\p,\f\in \F$) since annihilation operators are always
attached to negative powers of the formal variables.
\item \BF{Vacuum}\\
We choose the vacuum to be the zero mode state with no momentum and
without any creation operators, i.e.
\[ \1:=1\otimes\e^{\bf0} \]
so that \[
\V(\1,z)={\bf:}\e^{i{\bf0}\cdot{\bf Q}(z)}{\bf:}c_{\bf0}={\rm id}_\F \]
by the normalization condition (\ref{62a}).
\item \BF{Injectivity}\\
Observe that, when acting on the vacuum, terms involving only creation
operators survive in the expression for a vertex operator. Then it is
obvious that
\[ \p_{-1}\1=\Res{z}{z^{-1}\Vp\1}=\p\for{all }\p\in\F \]
In particular, $\Vp=0$ implies $\p=0$.
\item \BF{Conformal vector}\\
We claim that the element
\[ \w:=\frac12\sum_{\mu,\nu=1}^d{\bf e}_\mu(-1){\bf e}_\nu(-1)
   \eta^{\mu\nu}(\otimes\e^{\bf0}) \]
is a conformal vector of dimension $d$ and is independent of the choice
of the basis $\{{\bf e}_\mu\}$ of $\L$. We have
\baro \V(\w,z)&=&\frac12\sum_{\mu,\nu=1}^d{\bf:}{\bf e}_\mu(z)
                 {\bf e}_\nu(z){\bf:}\eta^{\mu\nu}\by{(\ref{65})} \\
              &=&\frac12\sum_{m,n\inZ}{\bf:}\alf_m\cdot\alf_n{\bf:}
                 z^{-m-n-2}\by{(\ref{63a})} \earo
(Note that in the last step we had to relie on nondegeneracy of the
lattice!) Thus
\[ \Ln\equiv\w_{n+1}=\frac12\sum_{m\inZ}{\bf:}\alf_m\cdot
\alf_{n-m}{\bf:} \]
in agreement with the well-known expression from string theory. Using
the oscillator commutation relations one indeed finds that the $\Ln$'s
obey (\ref{vi}).(see \cite{GSW88} for the calculation) \\
To establish the translation property of $\Lt$ we quickly get
\baro \Lt\e^\r&=&\r(-1)\e^\r\by{(\ref{63})} \\
      \Lt\r(-m)&=&m\r(-m-1)\quad\mbox{for }m>0\by{(\ref{60})} \earo
but on the other hand,
\baro \dz\V(\e^\r,z)&=&{\bf:}\r(z)\V(\e^\r,z){\bf:}=\V(\r(-1)\e^\r,z)
                       \by{(\ref{65}),(\ref{64})} \\
     \dz\V(\r(-m),z)&=&\frac1{(m-1)!}\left(\dz\right)^m\r(z)
                     =\V(m\r(-m-1),z)\by{(\ref{65}),(\ref{63a})} \earo
Together with derivation property of $\Lt$ and $\dz$ this proves
(\ref{tr}).\\
Finally, consider a typical homogeneous element of $\F$,
\[ \p=\bigg(\prod_{j=1}^N\s_j(-n_j)\bigg)\otimes\e^\r \]
Then
\baro \Lo\p&=&\bigg\{{\textstyle\frac12}\pp^2+\sum_{m\ge1}\alf_{-m}
              \cdot\alf_m\bigg\}
         \left(\bigg(\prod_{j=1}^N\s_j(-n_j)\bigg)\otimes\e^\r\right) \\
          &=&\bigg({\textstyle\frac12}\r^2+\sum_{j=1}^Nn_j\bigg)\p \earo
yields the desired grading of $\F$. Furthermore we observe that the
spectrum of $\Lo$ is nonegative and the eigenspaces of $\Lo$ are
finite-dimensional provided that $\L$ is a positive definite
lattice.\beenden
\subsection{Jacobi identity}
It is not surprising that by far the hardest axiom to prove is the
Jacobi identity because it contains most information about a vertex
algebra. We will not go much into details and will only mention the
important steps and crucial ideas.(see \cite{FLM88})\\
{\it Step 1:}\\
We make a change of variables. Let $\r_1,\ldots,\r_M,\s_1,\ldots,\s_N
\in\L$ and consider the formal sums
\baro R&\equiv&\prod_{i=1}^M\left(\e^{\sum_{m>0}\frac1m\r_i(-m)x_i^m}
                                \e^{\r_i}\right) \\
       &=&\prod_{i=1}^M\bigg(\sum_{m\ge0}p_m(\r_i(-1),\ldots,\r_i(-m))
          x_i^m\e^{\r_i}\bigg)\quad\in\F[\![x_1,\ldots,x_M]\!] \\
      S&\equiv&\prod_{j=1}^N\left(\e^{\sum_{n>0}\frac1n\s_j(-n)y_j^n}
                                \e^{\s_j}\right) \\
       &=&\prod_{j=1}^N\bigg(\sum_{n\ge0}p_n(\s_j(-1),\ldots,\s_j(-n))
          y_j^n\e^{\s_j}\Big)\quad\in\F[\![y_1,\ldots,y_N]\!] \earo
where the Schur polynomials $p_n(w_1,\ldots,w_n)$ are
\[ p_0=1,\quad p_1=w_1,\quad p_2=\frac1{2!}(w_1^2+w_2),\quad
   p_3=\frac1{3!}(w_1^3+3w_1w_2+2w_3),\quad\ldots \]
We note that the coefficients of the monomials in the formal variables
{\it span} $\F$ as $M$ and the $\r_i$'s and $N$ and the $\s_j$'s vary,
respectively. Hence it suffices to prove the Jacobi identity with $\p$
and $\f$ replaced by $R$ and $S$, respectively.\\
Using (\ref{64}) and (\ref{63}) we can immediately rewrite $R$ and $S$
as
\baro R&=&\ord\;\prod_{i=1}^M\V(\e^{\r_i},x_i)\;\ord\;\1 \\
 S&=&\ord\;\prod_{j=1}^N\V(\e^{\s_j},y_j)\;\ord\;\1 \earo
{\it Step 2:}\\
A lengthy but straightforward calculation which uses normal ordering
properties and (\ref{13}) shows that
\baro \V(R,z_1)&=&\ord\;\e^{\sum_{i=1}^M\sum_{m\ge1}\frac1{m!}\left(
      \frac{d}{dz_1}\right)^{m-1}\r_i(z_1)x_i^m}\V\bigg(\prod_{i=1}^M
      \e^{\r_i},z_1\bigg)\;\ord
  =\ord\prod_{i=1}^M\V(\e^{\r_i},z_1+x_i)\;\ord \\
      \V(S,z_2)&=&\ord\;\e^{\sum_{j=1}^N\sum_{n\ge1}\frac1{n!}\left(
      \frac{d}{dz_2}\right)^{n-1}\s_j(z_2)y_j^n}\V\bigg(\prod_{j=1}^N
      \e^{\s_j},z_2\bigg)\;\ord
  =\ord\prod_{j=1}^N\V(\e^{\s_j},z_2+y_j)\;\ord \earo
Hence
\[ \ord\;\V(R,z_1)\V(S,z_2)\;\ord=\prod_{1\le{i}\le{M} \atop 1\le{j}\le
   {N}}(-1)^{\r_i\cdot\s_j}\;\ord\;\V(S,z_2)\V(R,z_1)\;\ord \]
and
\baro \V(R,z_1)\V(S,z_2)&=&\prod_{1\le{i}\le{M} \atop 1\le{j}\le{N}}
                     \Big(z_1+(-z_2+x_i-y_j)\Big)^{\r_i\cdot\s_j}
                     \;\ord\;\V(R,z_1)\V(S,z_2)\;\ord \\
\V(S,z_2)\V(R,z_1)&=&\prod_{1\le{i}\le{M} \atop 1\le{j}\le{N}}
                     \Big(-z_2+(z_1+x_i-y_j)\Big)^{\r_i\cdot\s_j}
                     \;\ord\;\V(R,z_1)\V(S,z_2)\;\ord \earo
(all binomial expressions to be expanded in the second term!)\\
{\it Step 3:}\\
Fix $k\in\Z$ and a monomial $q=\prod_{i=1}^M\prod_{j=1}^Nx_i^{m_i}y_j^{n
_j}$, $m_i,n_j\ge0\ \forall{i,j}$.Choose $K\ge0$ such that $K+k\ge0$ and
$K+k\ge\deg q-\sum_{i=1}^M\sum_{j=1}^N\r_i\cdot\s_j$. Thus the
coefficient of $q$ and of each monomial of lower total degree than $q$
in
\baro F_K&\equiv&(z_1-z_2)^{K+k}
                 \prod_{1\le{i}\le{M} \atop 1\le{j}\le{N}}
                 \Big(z_1+(-z_2+x_i-y_j)\Big)^{\r_i\cdot\s_j} \\
         &=&(z_1-z_2)^{K+k}
                 \prod_{1\le{i}\le{M} \atop 1\le{j}\le{N}}
                 \sum_{m_i\ge0 \atop n_j\ge0}
            {\r_i\cdot\s_j \choose m_i}{\r_i\cdot\s_j-m_i \choose n_j}
    (-1)^{n_j}(z_1-z_2)^{\r_i\cdot\s_j-m_i-n_j}x_i^{m_i}y_j^{n_j} \earo
is a {\it polynomial} in $z_1-z_2$.\\
Let $V_q(z_1,z_2)$ denote the coefficient of $q$ in
\[ (z_1-z_2)^{K+k}\V(R,z_1)\V(S,z_2)=
                  F_K\;\ord\;\V(R,z_1)\V(S,z_2)\;\ord \]
Then the coefficient of $q$ in $(z_1-z_2)^k\V(R,z_1)\V(S,z_2)$ is
$(z_1-z_2)^{-K}V_q(z_1,z_2)$. Similarly we find that the coefficient of
$q$ in $(-z_2+z_1)^k\V(S,z_2)\V(R,z_1)$ is
$(-z_2+z_1)^{-K}V_q(z_1,z_2)$.\\
It follows that the coefficient of $q$ in
\[ (z_1-z_2)^k\V(R,z_1)\V(S,z_2)-(-z_2+z_1)^k\V(S,z_2)\V(R,z_1) \]
is, by (\ref{8}),
\[ -(-z_2)^{-K}\Theta\left(\Big(1-\frac{z_1}{z_2}\Big)^{-K}\right)
   V_q(z_1,z_2) \]
which is the coefficient of $z_0^{K-1}$ in
\[ \D{2}{1}{0}V_q(z_1,z_2)=\D{2}{1}{0}V_q(z_2+z_0,z_2) \]
by (\ref{8}),(\ref{6}) and (\ref{5}).\\
But $V_q(z_2+z_0,z_2)$ is the coefficient of $q$ in
\[ \ord\;\V(R,z_2+z_0)\V(S,z_2)\;\ord\;z_0^{K+k}
                 \prod_{1\le{i}\le{M} \atop 1\le{j}\le{N}}
                 \Big(z_0+(x_i-y_j)\Big)^{\r_i\cdot\s_j} \]
Hence
\( (z_1-z_2)^k\V(R,z_1)\V(S,z_2)-(-z_2+z_1)^k\V(S,z_2)\V(R,z_1) \)
is the coefficient of $z_0^{-k-1}$ in
\[ \D{2}{1}{0}\ord\;\V(R,z_2+z_0)\V(S,z_2)\;\ord\;
                 \prod_{1\le{i}\le{M} \atop 1\le{j}\le{N}}
                 \Big(z_0+(x_i-y_j)\Big)^{\r_i\cdot\s_j} \]
Note that the last expression is independent of $q$ and $K$! We conclude
that
\[ \lire{\D{0}{1}{2}\V(R,z_1)\V(S,z_2)\Di{0}{2}{1}\V(S,z_2)\V(R,z_1)}
        {=\D{2}{1}{0}\ord\;\V(R,z_2+z_0)\V(S,z_2)\;\ord\;
                 \prod_{1\le{i}\le{M} \atop 1\le{j}\le{N}}
                 \Big(z_0+(x_i-y_j)\Big)^{\r_i\cdot\s_j}} \]
{\it Step 4:}\\
On the other hand, using {\it Step 2},
\[ \V(R,z_0)S=\prod_{1\le{i}\le{M} \atop 1\le{j}\le{N}}
              \Big(z_0+(x_i-y_j)\Big)^{\r_i\cdot\s_j}
              \;\ord\;\prod_{i=1}^M\V(\e^{\r_i},z_0+x_i)
              \prod_{j=1}^N\V(\e^{\s_j},y_j)\;\ord\;\1 \]
Thus
\baro \V(\V(R,z_0)S,z_2)
           &=&\prod_{1\le{i}\le{M} \atop 1\le{j}\le{N}}
              \Big(z_0+(x_i-y_j)\Big)^{\r_i\cdot\s_j}
              \;\ord\;\prod_{i=1}^M\V(\e^{\r_i},z_2+z_0+x_i)
              \prod_{j=1}^N\V(\e^{\s_j},z_2+y_j)\;\ord\ \\
           &=&\ord\;\V(R,z_2+z_0)\V(S,z_2)\;\ord\;
              \prod_{1\le{i}\le{M} \atop 1\le{j}\le{N}}
              \Big(z_0+(x_i-y_j)\Big)^{\r_i\cdot\s_j} \earo
This completes the proof of the Jacobi identity.
\subsection{Example 1: Affine Lie algebras}
Suppose that $\L$ is a positive definite even lattice. Obviously,
$\Fo{0}$ is one-dimensional and the spectrum of $\Lo$ is nonegative
so that $\Fo{1}$ is a Lie algebra. Its elements are easy to describe,
\[ \Fo{1}=\{\e^\r|\r\in\L_2\}\oplus\{\s(-1)|\s\in\L\} \]
where $\L_2\equiv\{\r\in\L|\r^2=2\}$ denotes the set of lattice vectors
of squared length two. We can work out the antisymmetric product and the
invariant bilinear form explicitly:
\baro [\r(-1),\s(-1)]&=&\r(-1)_0\s(-1)=\Res{z}{\r(z)\big(\s(-1)\big)} \\
          &=&\Res{z}{\sum_{m\inZ}z^{-m-1}\r(m)\big(\s(-1)\big)} \\
          &=&0\by{(\ref{60})} \\[1em]
      [\r(-1),\e^\s]&=&\r(-1)_0\e^\s=\Res{z}{\r(z)\big(\e^\s\big)} \\
          &=&\Res{z}{\sum_{m\inZ}z^{-m-1}\r(m)\big(\e^\s\big)} \\
          &=&(\r\cdot\s)\e^\s\by{(\ref{63})} \\[1em]
      [\e^\r,\e^\s]&=&\e^\r_0\e^\s=
                             \Res{z}{\e^{\int\r_<(z)dz}\e^\r z^{\r(0)}
                             \e^{\int\r_>(z)dz}\big(\e^\s\big)} \\
   &=&\Res{z}{\e^{\sum_{m>0}\frac1m\r(-m)z^m}z^{\r\cdot\s}\e^\r\e^\s} \\
&=&\cases{0  & if $\r\cdot\s\ge0$ \cr
          \epsilon(\r,\s)\e^{\r+\s}  & if $\r\cdot\s=-1$ \cr
          \r(-1)  & if $\r\cdot\s=-2$} \earo
We note that the Schwarz inequality yields $|\r\cdot\s|\le2$. Moreover
$\r\cdot\s=-1\iff\r+\s\in\L_2$ and $\r\cdot\s=-2\iff\r+\s=0$ for $\r,\s
\in\L_2$.\\
Similarly, we find
\baro \lx\r(-1),\s(-1)\rx&=&\r(-1)_1\s(-1)=\r\cdot\s \\
      \lx\r(-1),\e^\s\rx&=&\r(-1)_1\e^\s=0 \\
      \lx\e^\r,\e^\s\rx&=&\e^\r_1\e^\s=
                                \cases{0  & if $\r\cdot\s\ge-1$ \cr
                                       1  & if $\r\cdot\s=-2$} \earo
Thus we have arrived at a root space decomposition of the Lie algebra
$\Fo{1}$ where the root lattice is precisely the lattice $\L$ and the
set of roots is given by $\L_2$. \\
For the affine Lie algebra $\hat{\Fo{1}}$ we find the formulas
\baro [\r(-1)_m,\s(-1)_n]&=&m(\r\cdot\s)\delta_{m+n,0} \\ {}
      [\r(-1)_m,\e^\s_n]&=&(\r\cdot\s)\e^\s_{m+n} \\ {}
      [\e^\r_m,\e^\s_n]&=&\cases{0  & if $\r\cdot\s\ge0$ \cr
              \epsilon(\r,\s)\e^{\r+\s}_{m+n}  & if $\r\cdot\s=-1$ \cr
       \r(-1)_{m+n}+m\delta_{m+n,0}  & if $\r\cdot\s=-2$} \earo
The first equation is no surprise since the operators $\r(-1)_m$ are
nothing but the oscillators we have started with,
\[ \r(-1)_m=\Res{z}{z^m\r(z)}=\r(m) \]
while the operators $\e^\r_m$ are those occuring in the Frenkel-Kac
construction \cite{FreKac80} of affine Lie algebras,
\[ \e^\r_m=\Res{z}{z^m\;\ord\;\e^{i\r\cdot{\bf Q}(z)}\;\ord\;}c_\r \]
In physics literature this construction of affine Lie algebras is prese%
nted pedagogically in \cite{Slan88},\cite{KMPS90},\cite{GodOli85},\cite%
{GodOli86},\cite{Godd86}.
\subsection{Example 2: Fake Monster Lie algebra}
Things become more complicated when we move away from the lattice $\L$
being Euclidian. Let us consider the unique 26-dimensional even
unimodular Lorentzian lattice $\II$, which can be taken to be the
lattice of all points ${\bf x}=(x_1,\ldots,x_{25}|x_0)$ for which the
$x_\mu$ are all in $\Z$ or all in $\Z+\frac12$ and which have integer
inner product with the vector ${\bf l}=(\frac12,\ldots,\frac12|\frac12)
$, all norms and inner products being evaluated in the Lorentzian metric
${\bf x}^2=x_1^2+\cdots+x_{25}^2-x_0^2$. (cf. \cite{Serr73}) In physics
this corresponds to an open bosonic string moving in 26-dimensional
spacetime compactified on a torus so that the momenta lie on a lattice.
\cite{LeScheWa89} Calculations in connection with the automorphism
group of $\II$ show that a set of positive norm simple roots for $\II$
is given by the subset of vectors $\r$ in $\II$ for which $\r^2=2$ and
$\r\cdot\Weyl=-1$ where the Weyl vector is
$\Weyl=(0,1,2,\ldots,24|70)$.(cf. \cite{CoSl82b},\cite{Con83}) In fact,
these simple roots generate the reflection group of $\II$, where the
reflection $\sigma_\r$ associated with a root $\r$ is defined as
$\sigma_\r({\bf x})={\bf x}-\frac2{\r^2}({\bf x}\cdot\r)\r$. We shall
also call the positive norm simple roots of $\II$ Leech roots since
Conway has shown that this subset is indeed isometric to the Leech
lattice, the unique 24-dimensional even unimodular Euclidian lattice
with no vectors of square length two. For further informations about
the Leech lattice and the other Niemeier lattices the reader may wish
to consult \cite{Niem73},\cite{CoPaSl82},\cite{CoSl82a},\cite{Borc85}.

We now define a Kac-Moody algebra ${\rm L}_\infty$, of infinite
dimension and rank, as follows (see \cite{BoCoQuSl84}) : ${\rm
L}_\infty$ has three generators $e(\r)$, $f(\r)$, $h(\r)$ for each
Leech root $\r$, and is presented by the following relations,
\baro [h(\r),e(\s)]&=&(\r\cdot\s)e(\s) \\ {}
      [h(\r),f(\s)]&=&-(\r\cdot\s)f(\s) \\  {}
      [e(\r),f(\r)]&=&h(\r) \\ {}
      [e(\r),f(\s)]&=&0 \\  {}
      [h(\r),h(\s)]&=&0 \\
      \Big({\rm\,ad}e(\s)\Big)^{1-\r\cdot\s}e(\r)&=&
      \Big({\rm\,ad}f(\s)\Big)^{1-\r\cdot\s}f(\r)=0  \earo
where $\r$ and $\s$ are distinct Leech roots. In a sort of Dynkin
diagram for ${\rm L}_\infty$ two nodes $\r$,$\s$ are joined by
$-\r\cdot\s$ lines and a portion of the (infinite) graph looks like the
following figure (cf. \cite{Con83}):
\[ \beginpicture
\setcoordinatesystem units <.5mm,.5mm>
\circulararc 360 degrees from 100 200 center at 100 100
\setquadratic \plot 112.5 .5 126 108 52 187.5 /
              \plot 198 81 100 127.5 2 81 /
              \plot 148 187.5 74 108 87.5 .5 /
              \plot 31 172 84 77.5 190.5 58 /
              \plot 9.5 58 116.5 77.5 169 172 /
              \plot 147 35 149.5 115.5 100 180 /
              \plot 100 180 50.5 115.5 53 35 /
              \plot 53 35 130 58.5 176 124 /
              \plot 176 124 100 150.8 24 124 /
              \plot 24 124 70 58.5 147 35 /
\setlinear \plot 100 200     
                 100 127.5 / 
           \plot 195 130     
                 126 108   / 
           \plot 158.5 19    
                 116.5 77.5 /
           \plot 41.5 19     
                 84 77.5   / 
           \plot 5 130       
                 74 108    / 
           \plot 100 127.5   
                 126 108     
                 116.5 77.5  
                 84 77.5     
                 74 108      
                 100 127.5 /
\put {\ball} <\X,\Y> at 100 200     
\put {\ball} <\X,\Y> at 124 197     
\put {\ball} <\X,\Y> at 148 187.5   
\put {\ball} <\X,\Y> at 169 172     
\put {\ball} <\X,\Y> at 184.5 153   
\put {\ball} <\X,\Y> at 195 130     
\put {\ball} <\X,\Y> at 199.5 106   
\put {\ball} <\X,\Y> at 198 81      
\put {\ball} <\X,\Y> at 190.5 58    
\put {\ball} <\X,\Y> at 177 36      
\put {\ball} <\X,\Y> at 158.5 19    
\put {\ball} <\X,\Y> at 137 7       
\put {\ball} <\X,\Y> at 112.5 .5    
\put {\ball} <\X,\Y> at 76 197      
\put {\ball} <\X,\Y> at 52 187.5    
\put {\ball} <\X,\Y> at 31 172      
\put {\ball} <\X,\Y> at 15.5 153    
\put {\ball} <\X,\Y> at 5 130       
\put {\ball} <\X,\Y> at .5 106      
\put {\ball} <\X,\Y> at 2 81        
\put {\ball} <\X,\Y> at 9.5 58      
\put {\ball} <\X,\Y> at 23 36       
\put {\ball} <\X,\Y> at 41.5 19     
\put {\ball} <\X,\Y> at 63 7        
\put {\ball} <\X,\Y> at 87.5 .5     
\put {\ball} <\X,\Y> at 100 127.5   
\put {\ball} <\X,\Y> at 126 108     
\put {\ball} <\X,\Y> at 116.5 77.5  
\put {\ball} <\X,\Y> at 84 77.5     
\put {\ball} <\X,\Y> at 74 108      
\put {\ball} <\X,\Y> at 100 180     
\put {\ball} <\X,\Y> at 176 124     
\put {\ball} <\X,\Y> at 147 35      
\put {\ball} <\X,\Y> at 53 35       
\put {\ball} <\X,\Y> at 24 124      
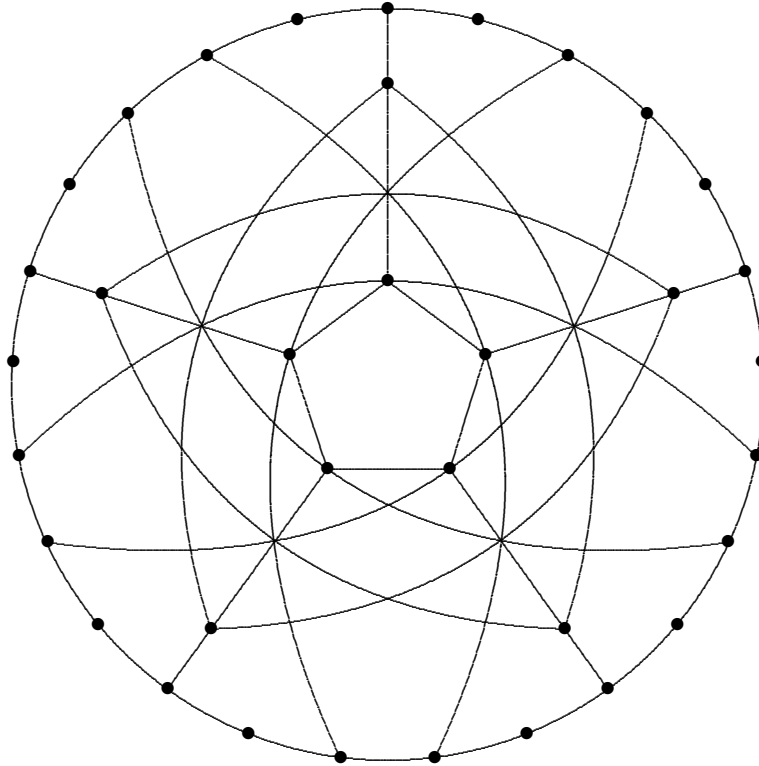
\put {Figure 1: Part of the Dynkin diagram of ${\rm L}_\infty$}
     at 100 -10
\endpicture \]
Besides infiniteness another fascinating feature of this graph is the
obvious fivefold symmetry. We have checked the next (w.r.t. increasing
value of the time coordinate) fifteen Leech roots and found that they
also fit nicely into this pentagram-like symmetry. If this pattern
turned out to be true for the whole diagram then it should be somehow
reflected in the structure of the automorphism group for the Leech
lattice.

Frenkel \cite{Fren85} proved that the dimension of the root space
corresponding to any root $\r$ of ${\rm L}_\infty$ is {\it at most}
$p_{24}(1-\frac12\r^2)$ where $p_{24}(n)$ is the number of partitions
of $n\in\N$ into parts of 24 colours, i.e.
\[ \Delta(q)^{-1}\equiv\sum_{n\ge0}p_{24}(1+n)q^n=q^{-1}\prod_{n>0}
   (1-q^n)^{-24}=q^{-1}+24+324q+3200q^2+\cdots \]

Let us return to the vertex algebra \vertex associated with the
lattice $\II$. It is easy to see that, for any Leech root $\r$, the
elements $\e^\r$, $\e^{-\r}$, $\r(-1)$ describe the physical states
of conformal weight one, i.e. they lie in $\Ph{1}$. We can immediately
infer from Example 1 that ${\rm L}_\infty$ is mapped into the
Lie algebra $\fake:=\Ph{1}/{\rm kernel}(\_,\_)$ by
\baro e(\r)&\mapsto&\e^\r \\
      f(\r)&\mapsto&\e^{-\r} \\
      h(\r)&\mapsto&\r(-1) \earo
Hence ${\rm L}_\infty$ is a subalgebra of $\fake$. Note that
we have to divide out the kernel of $(\_,\_)$ since in 26 dimensions
additional null physical states besides $\Lt\Fo{0}\cap\Ph{1}$ occur.

However this is not the whole story about $\fake$ since
${\rm L}_\infty$ is a {\it proper} subalgebra of $\fake$.
To see this we exploit the connection with string theory as presented
in \cite{GSW88}. So far we have only considered the tachyonic ground
states $\e^\r$ and the oscillators $\r(-1)_m\equiv\r(m)$. Hence we
still have to add the DDF operators which correspond to photon
emission vertices. In fact, the elements $\ix(-1)\e^{m\weyl}$ for
$m\in\Z$, $\ix\in\II$, are physical states provided $\ix\cdot\Weyl=0$,
i.e. provided that $\ix$ is a transverse polarization vector. $\ix$
must not be proportional to the (lightlike!) Weyl vector $\Weyl$
because otherwise the corresponding state would be a null physical
state (more precisely, it would lie in $\Lt\Fo{1}\cap\Ph{1}$). This
leaves us with 24 degrees of freedom for the polarization. Additionally
we have to incorporate the states $m\Weyl(-1)$ for $m\in\Z$. The
calculation of the Lie bracket gives
\baro [\r(-1),\ix(-1)\e^{m\weyl}]&=&m(\r\cdot\Weyl)\ix(-1)\e^{m\weyl}
                                    \qquad\forall\r\in\II \\ {}
    [\e^\r,\ix(-1)\e^{m\cdot\weyl}]&=&0\If{}m(\r\cdot\Weyl)\ge1 \\ {}
      [\ix(-1)\e^{m\weyl},\ita(-1)\e^{n\weyl}]&=&m(\ix\cdot\ita)\Weyl(
                                    -1)\delta_{m+n,0} \earo
where we used $\ix\cdot\Weyl=\ita\cdot\Weyl=\Weyl\cdot\Weyl=0$ and
$\Weyl(-1)\e^{m\weyl}\e^{n\weyl}=\frac1{m+n}\Lt(\e^{m\weyl}\e^{n\weyl})
\in\Lt\Fo{0}$ if $m+n\neq0$.

The no-ghost theorem together with Borcherds' theorem \cite{Borc90}
tell us that this is a complete set of generators
for the spectrum of physical states and that the dimension of a subspace
of momentum ${\bf x}\in\II$ is {\it exactly} given by $p_{24}(1-\frac12
{\bf x}^2)$. Moreover, the bilinear form $(\_,\_)$ is positive
definite on any subspace of nonzero momentum $\bf x$.

Let us summarize: We define the {\bf fake Monster Lie algebra} $\fake$
to be the Lie algebra with root lattice $\II$, whose simple roots
are the simple roots of the Kac-Moody algebra ${\rm L}_\infty$, together
with the positive integer multiples of the Weyl vector $\Weyl$, each
with multiplicity 24. Then any nonzero root ${\bf x}\in\II$ of $\fake$
has multiplicity $p_{24}(1-\frac12{\bf x}^2)$. Note that the set of
simple roots is characterized by the condition $\r\cdot\Weyl=-\frac12\r
^2$. The Lie algebra $\fake$ was first defined by Borcherds in
\cite{Borc90}.

This result is quite astonishing because the fake Monster Lie algebra
is {\it not} a Kac-Moody algebra due to the presence of the lightlike
simple Weyl roots which violate an axiom for these algebras.(cf. \cite%
{Mood79},\cite{Kac90}) Nevertheless, the structure of $\fake$ resembles
a Kac-Moody algebra very well. It was Borcherds' great achievement to
observe that one can define generalized Kac-Moody algebras by allowing
imaginary ($\equiv$ nonpositive norm) simple roots in the defining
axioms of Kac-Moody algebras \cite{Borc91}. To complete our first
example of a Borcherds algebra we introduce a set of 24 orthonormal
transverse polarization vectors $\ix_i$, i.e. $\ix_i\cdot\Weyl=0$,
$\ix_i\cdot\ix_j=\delta_{ij}$ for $1\le i,j\le24$, and consider the
following generators:
\baro \ix_i(-1)\e^{m\weyl}&\mapsto&e_i(m\Weyl) \\
      \ix_i(-1)\e^{-m\weyl}&\mapsto&f_i(m\Weyl) \\
                 m\Weyl(-1)&\mapsto&h_i(m\Weyl) \earo
for $1\le i\le24$ and $m>0$. In addition to the relations for ${\rm L}_
\infty$ we get
\baro [e(\r),f_i(m\Weyl)]&=&0 \\ {}
      [f(\r),e_i(m\Weyl)]&=&0 \\ {}
    [e_i(m\Weyl),f_j(n\Weyl)]&=&\delta_{mn}\delta_{ij}h_i(m\Weyl) \\ {}
      [e_i(m\Weyl),e_j(n\Weyl)]&=&0 \\ {}
      [f_i(m\Weyl),f_j(n\Weyl)]&=&0 \\ {}
      [h(\r),e_i(m\Weyl)]&=&-me_i(m\Weyl) \\ {}
      [h(\r),f_i(m\Weyl)]&=&mf_i(m\Weyl) \\ {}
      [h_i(m\Weyl),e(\r)]&=&-me(\r) \\ {}
      [h_i(m\Weyl),f(\r)]&=&mf(\r) \\ {}
      [h_i(m\Weyl),e_j(n\Weyl)]&=&0 \\ {}
      [h_i(m\Weyl),f_j(n\Weyl)]&=&0 \\ {}
      [h(\r),h_i(m\Weyl)]&=&0 \\ {}
      [h_i(m\Weyl),h_j(n\Weyl)]&=&0 \earo
The Cartan matrix looks as following:
\[ \left( \begin{array}{rr|rrr}
   &&\multicolumn{1}{c}{\vdots}&\multicolumn{1}{c}{\vdots}&  \\
   \raisebox{3mm}[5mm]{\makebox[0pt][l]{\Large${\rm L}_{(\infty)}$}}
    & & -1^{(24)}&-2^{(24)}&\ldots\\ \hline
   \ldots&(-1^{(24)})^T&0_{24}&0_{24}&\ldots\\
   \ldots&(-2^{(24)})^T&0_{24}&0_{24}&\ldots\\
   &\multicolumn{1}{c|}{\vdots}&\multicolumn{1}{c}{\vdots}&
    \multicolumn{1}{c}{\vdots}&\ddots \end{array} \right) \]
where $0_{24}$ is the $24\times24$ zero matrix, the
24 dimensional row vectors $-m^{(24)}$,$m\ge0$, are given by
\[ -m^{(24)}\equiv\left( \begin{array}{ccc}
   -m&\ldots&-m \end{array}\right) \]
and ${\rm L}_{(\infty)}$ denotes the infinite-dimensional Cartan matrix
for the Leech roots, i.e. with entry $\r\cdot\s$ in the $\r$th row and
$\s$th column for Leech roots $\r,\s$.
\section{Borcherds algebras $\equiv$ generalized Kac-Moody algebras}
\subsection{Definition and properties} \labelx{5.1}
The purpose of this section is to develop the formal aspects of
Borcherds algebras as presented in \cite{Borc88},\cite{Borc91},\cite%
{Borc92}. Borcherds algebras are also mentioned in \cite{Kac90}.
\begin{defi} {\bf:} \labelx{d3} \\
Let $A=(a_{ij})$, $i,j\in I$, be a symmetric real matrix with no zero
columns, possibly infinite but countable, satisfying the following
properties:
\beginnen \item[(i)] either $a_{ii}=2$ or $a_{ii}\le0$
          \item[(ii)] $a_{ij}\le0$ if $i\neq j$
          \item[(iii)] $a_{ij}\in\Z$ if $a_{ii}=2$ \beenden
Then the {\bf universal generalized Kac-Moody algebra} associated to
$A$ is defined to be the Lie algebra $\ga$ given by the following
generators and relations.\\
Generators: Elements $e_i,\ f_i,\ h_{ij}$ for $i,j\in I$ \\
Relations:
\beginnen \item[(1)] $[e_i,f_j]=h_{ij}$
          \item[(2)] $[h_{ij},e_k]=\delta_{ij}a_{jk}e_k,\
                   [h_{ij},f_k]=-\delta_{ij}a_{jk}f_k$
          \item[(3)] $({\rm ad}\,e_i)^{1-a_{ij}}e_j=0,\quad
                  ({\rm ad}\,f_i)^{1-a_{ij}}f_j=0
                  \If{}a_{ii}=2\mbox{ and }i\neq j$
         \item[(4)] $[e_i,e_j]=0,\ [f_i,f_j]=0\If{}a_{ii}\le0,a_{jj}\le0
                 \mbox{ and }a_{ij}=0$ \beenden
\end{defi}
Let us make some remarks and list important properties of universal
generalized Kac-Moody algebras. \beginnen
  \item There is a unique invariant bilinear form $(\_,\_)$ on $\ga$
        such that $(e_i,f_j)=\delta_{ij}$; invariance and relations (1),
         (2) then imply $(h_{ii},h_{jj})=a_{ij}$.
  \item If $a_{ii}=2$ for all $i\in I$ then $\ga$ is the same as the
        ordinary Kac-Moody algebra with symmetrized Cartan matrix $A$.
        In general, $\ga$ has almost all the properties that ordinary
        Kac-Moody algebras have, and the only major difference is
        that generalized Kac-Moody algebras are allowed to have
        imaginary simple roots.
  \item The Jacobi identity applied to the elements $h_{ij}$, $e_k$,
        $f_l$ yields
        \[ [h_{ij},h_{kl}]=\delta_{ij}(a_{jk}-a_{jl})h_{kl} \]
        so that \beginnen
        \item $h_{ij}$ lies in the centre of $\ga$ if $i\neq j$,
        \item all the $h$'s commute with each other,
        \item $h_{ij}=0$ if the $i$th and the $j$th columns of $A$ are
              not equal \beenden
        The elements $h_{ij}$ for which the $i$th and the $j$th columns
        of $A$ are equal form a basis for an abelian subalgebra of
        $\ga$, called its {\bf Cartan subalgebra} $\cart$. In the case
        of ordinary Kac-Moody algebras, the $i$th and the $j$th columns
        of $A$ cannot be equal unless $i=j$, so the only nonzero
        elements $h_{ij}$ are those of the form $h_{ii}$ which are
        usually denoted by $h_i$. The reason why we need the elements
        $h_{ij}$ for $i\neq j$ is that $\ga$, so defined, is equal to
        its own universal central extension.
  \item We can define a $\Z$-gradation of $\ga$ by $\deg(e_i)=-
        \deg(f_i)=n_i$ where $\{n_i|i\in I\}$ is a collection of
        positive integers with finite repetitions. The degree zero
        piece of $\ga$ is the Cartan subalgebra $\cart$.
  \item $\ga$ has an antilinear involution $\theta$ with $\theta(e_i)=
        -f_i$, $\theta(f_i)=-e_i$, $\theta(h_{ij})=-h_{ji}$, called the
        {\bf Cartan involution}.
  \item The contravariant form $(x,y)_0:=(\theta(x),y)$ is "almost
        positive definite" on $\ga$ which means that $(x,x)_0>0$
        whenever $x$ is a homogeneous element of nonzero degree in
        $\ga$.
  \item The {\bf root lattice} $\L_R$ is defined to be the free abelian
        group generated by elements $\r_i$ for $i\in I$, with the
        bilinear form given by $\r_i\cdot\r_j:=a_{ij}$. The elements
        $\r_i$ are called the {\bf simple roots}. The universal
        generalized Kac-Moody algebra $\ga$ is $\L_R$-graded by letting
        $\cart$ have degree zero, $e_i$ have degree $\r_i$ and $f_i$
        have degree $-\r_i$. The root space of an element $\r\in\L_R$
        is the vector space of elements of $\ga$ of that degree; if
        $\r$ is nonzero and has a nonzero root space then $\r$ is called
        a {\bf root} of $\ga$. A root $\r$ is called {\bf positive} if
        it is a sum of simple roots, and {\bf negative} if $-\r$ is
        positive. Every root is either positive or negative. A root $\r$
        is called {\bf real} if $\r^2>0$ and {\bf imaginary} otherwise.
       In \cite{Kac90} there are proved a number of facts about the root
        systems of universal generalized Kac-Moody algebras.
  \item There is a {\bf denominator formula} for universal generalized
        Kac-Moody algebras. This states that
        \[ \e^{\weyl}\prod_{\r>0}(1-\e^\r)^{{\rm mult}(\r)}=
          \sum_{w\in{\cal W}}\det(w)w\left(\e^{\weyl}\sum_\r\epsilon(\r)
           \e^\r\right) \]
        Here $\Weyl$ is the Weyl vector ($\equiv$ vector with
        $\Weyl\cdot\r=-\frac12\r^2$ for all simple roots), $\r>0$ means
        that $\r$ is a positive root, $\cal W$ is the Weyl group
        ($\equiv$ group of isometries of $\L_R$ generated by the
        reflections $\sigma_i(\r):=\r-(\r\cdot\r_i)\r_i$ corresponding
        to the {\it real} simple roots $\r_i$); $\det(w)$ is defined
        to be $+1$ or $-1$ depending on whether $w$ is the product of
        an even or odd number of reflections and $\epsilon(\r)$ is
        $(-1)^n$ if $\r$ is the sum of $n$ distinct pairwise orthogonal
        imaginary simple roots, and zero otherwise. \\
        Note that the Weyl vector $\Weyl$ may be replaced by any vector
        having inner product $-\frac12\r^2$ with all {\it real} simple
        roots $\r$ since $\e^{w(\weyl)-\weyl}$ only involves inner
        products of $\Weyl$ with the real simple roots. \\
        For ordinary Kac-Moody algebras there are no imaginary simple
        roots , so the sum over $\r$ equals one and we end up with the
        well-known denominator formula.
  \item There is a natural homomorphism of abelian groups from the root
        lattice $\L_R$ to the Cartan subalgebra $\cart$ taking $\r_i$
        to $h_{ii}\equiv h_i$ which preserves the bilinear form. This
        map is not usually injective. It is possible for $n$ imaginary
        simple roots to have the same image $h_i$ in $\cart$ in which
        case we say, by abuse of language, that $\r_i$ is a simple root
        "of multiplicity $n$". \beenden

If we take the quotient of a universal generalized Kac-Moody algebra
we obtain a generalized Kav-Moody algebra.
\begin{defi} {\bf:}\labelx{d4} \\
A {\bf generalized Kac-Moody algebra} is a Lie algebra $\G$ with an
almost positive definite contravariant form, which means that $\G$ has
the following properties: \beginnen
 \item[(1)] $\G=\bigoplus_{i\inZ}\G_i$ is $\Z$-graded with
            $\dim\G_i<\infty$
 \item[(2)] $\G$ has an involution $\theta$ which acts as $-1$
            on $\G_0$ and maps $\G_i$ to $\G_{-i}$
 \item[(3)] $\G$ carries an invariant bilinear form $(\_,\_)$
            invariant under $\theta$ such that $(\G_i,\G_j)=0$ unless
            $i+j=0$
 \item[(4)] The contravariant form $(x,y)_0:=-(\theta(x),y)$ is
            positive definite on $\G_i$ if $i\neq0$
 \item[(5)] $\G_0\subset[\G,\G]$ \beenden \end{defi}
(If the last condition is omitted we may add an abelian algebra of outer
derivations to a generalized Kac-Moody algebra. If the $i$th and the
$j$th column of $A$ are equal then $\ga$ has an outer derivation $d$
defined by $[d,e_i]=e_j$, $[d,e_j]=-e_i$, $[d,f_i]=f_j$, $[d,f_j]=-f_i$,
and $[d,e_k]=[d,f_k]=0$ if $k\neq i,j$. These outer derivations do not
always commute with the elements of the Cartan subalgebra $\cart$.)

The main theorem about generalized Kac-Moody algebras states that we
can construct any generalized Kac-Moody algebra from some universal
generalized Kac-Moody algebra by factoring out some of the centre and
adding a commuting algebra of outer derivations.

It is easy to see that the fake Monster Lie algebra $\fake$ is a
generalized Kac-Moody algebra in the sense of Definition \ref{d4}.
If we fix any negative norm vector $\bf x$ of $\II$ not perpendicular to
any Leech root then we can make $\fake$ into a $\Z$-graded Lie algebra
by using the inner product with $\bf x$ as the degree. Then all the
conditions of the definition of a generalized Kac-Moody algebra are
satisfied for $\fake$.
\subsection{Simple examples}
Let us consider the two simplest examples of Borcherds algebras, namely,
the Borcherds extensions of $\su(2)$ and $\widehat{\su(2)}$
with one lightlike simple root.(see \cite{Slan91})

We start with the following generalized Cartan matrix:
\[ A=\left(\begin{array}{rr} 0&-1\\ -1&2 \end{array}\right) \]
The simple roots $\r_1$ and $\r_2$ are imaginary and real, respectively.
The scalar product of two roots $\r=m_1\r_1+m_2\r_2$, $\s=n_1\r_1+n_2
\r_2$ is given by $\r\cdot\s=-m_1n_2-m_2n_1+2m_2n_2$. A presentation of
the corresponding Borcherds algebra $\ga$ is given in terms of six
generators $e_1$, $e_2$, $f_1$, $f_2$, $h_1$, $h_2$, and relations,
\baro [e_i,f_j]&=&\delta_{ij}h_i \\ {}
      [h_i,e_j]&=&a_{ij}e_j \\ {}
      [h_i,f_j]&=&-a_{ij}f_j \\ {}
      [e_2,[e_2,e_1]]&=&0 \\ {}
      [f_2,[f_2,f_1]]&=&0 \earo
We define the {\bf fundamental dominant weights} $\l_1$,$\l_2$ relative
to the simple roots $\r_1$,$\r_2$ by the dual base condition
\[ \l_i\cdot\r_j\stackrel!=\delta_{ij}\qquad i,j=1,2 \]
so that $\l_1=-2\r_1-\r_2$ and $\l_2=-\r_1$. An important result of
representation theory (see \cite{Hump72},\cite{Vara74},\cite{FulHar91},
e.g.) then tells us that any highest weight associated with a highest
weight representation can be written as a sum of fundamental dominant
weights.\\
To actually compute the weight multiplicities for a highest weight
representation of a Borcherds algebra one derives from the denominator
formula some useful recursion formulas which can be put on a computer.
A part of the result for the $(1,0)$ fundamental representation (i.e.
with $\l_1$ as highest weight) of $\ga$ is listed in the following
table:
\[ \begin{array}{|c|l|l|l|} \hline
            & \mbox{multiplicity of weight} & &  \\
   \rb{$n_1$} & \l=n_1\r_1+[n_2]\r_2 & \rb{$\su(2)$ content} &
   \rb{as tensor products} \\ \hline
   0 & [0] & (0) & (0) \\
   1 & [1]+1[0] & (1) & (1) \\
   2 & [2]+2[1]+1[0] & (2)+1(0) & (1)\otimes(1) \\
   3 & [3]+3[2]+3[1]+1[0] & (3)+2(1) & (1)\otimes(1)\otimes(1) \\
   4 & [4]+4[3]+6[2]+4[1]+1[0] & (4)+3(2)+2(0) &
            (1)\otimes(1)\otimes(1)\otimes(1) \\
   \vdots & \vdots & \vdots & \vdots \\ \hline \end{array} \]
Note that we slice the representation with the imaginary simple root
$\r_1$ which means that we regard $n_1$ as a number operator eigenvalue.
At each level $n_1$ we can rewrite the portion of the representation
in terms of $\su(2)$ representations with highest weight
$\mbox{\boldmath$\Lambda$}(h_2)=2l$. For example, at slice
$n_1=3$ we find weights $\l=n_1\r_1+n_2\r_2$ with $n_2=0,1,2,3$ and
multiplicity $1,3,3,1$,respectively. Thus we have one four-dimensional
(iso)spin $\frac32$ and two two-dimensional (iso)spin $\frac12$
representations indicated by $2(1)+1(3)$ in the table. And this is
nothing but the tensor product of three (iso)spin $\frac12$.(see also
the review \cite{Slan81}) \\
We observe that the multiplicity of the weight $\l=n_1\r_1+n_2\r_2$
equals ${n_1 \choose n_2}$, the total number of states at slice $n_1$
being $\sum_{n_2=0}^{n_1}{n_1 \choose n_2}=2^{n_1}$ which is the
coefficient of $q^{n_1}$ in the partition function $P(q)=\frac1{1-2q}$.
This shows that the $\su(2)$ structure at slice $n_1$ is the tensor
product of the two-dimensional (iso)spin $\frac12$ representation with
itself $n_1$ times with no symmetry or antisymmetry constraints.\\
The number operator, $N$, and the diagonalized operator of $\su(2)$
, $I_3$, can be expressed in terms of the $\{h_1,h_2\}$ basis of the
Cartan subalgebra as $N=-2h_1-h_2$ and $2I_3=h_2$, respectively, which
shows that the operator $N$ counting the number of $e_1$ operators lies
in the Cartan subalgebra and corresponds to the root $\r_N=\l_1=-2\r_1-
\r_2$.

In the case of extending the affine Lie algebra $\widehat{\su(2)}$ we
consider the following Cartan matrix:
\[ A=\left(\begin{array}{rrr}
                    0&-1&0\\ -1&2&-2\\ 0&-2&2 \end{array}\right) \]
The roots are of the form $\r=m_1\r_1+m_2\r_2+m_3\r_3$. It is most
natural to look break the $(1,0,0)$ representation of the corresponding
Borcherds algebra $\ga$ into representations of $\widehat{\su(2)}$
,i.e. again we slice the representation with the imaginary simple root
$\r_1$. Computer calculations for the first values of $n_1$ show that
the $\widehat{\su(2)}$ structure at slice $n_1$ is the tensor
product of the $(1,0)$ fundamental representation of $\widehat{\su(2)}$
with itself $n_1$ times. In other words, starting from
two-dimensional current algebra, the fundamental $(1,0,0)$
representation of the simplest Borcherds extension contains a vacuum
at $n_1=0$, single particles at $n_1=1$, two particle states at $n_2=2$
, and so on. The surprising result is that the full multiparticle space
of states is included in this single representation. Moreover both
number operator and Hamilton operator are members of the Cartan
subalgebra: $N=-h_2-h_3$, $\Lo=-h_1-h_2-\frac12h_3$, $2I_3=h_2$.
%

To give an interpretation and a possible application of this feature
we make a short digression to the question of symmetry in quantum
theory. We will follow \cite{BarRac86} and \cite{Wybo74}. \\[1em]
Consider the differential operator $W=i\frac\partial{\partial t}-H$
where $H$ denotes the Hamilton operator of some nonrelativistic quantum
system. We define wavefunctions $\p$ of the system as solutions of the
Schr\"odinger equation $W\p=0$. If there are operators $G_j$, $j=1,
\ldots,n$, forming a Lie algebra $\G$ and satisfying, on the space
$\Phi$ of solutions, $[W,G_j]\p=0$ then $\Phi$ is a representation
space for the {\bf dynamical Lie algebra} $\G$ of the quantum system.
In general, $\G$ contains time-dependent operators $G(t)$ satisfying
the Heisenberg equation
\[ [i\frac\partial{\partial t},G(t)]\p=[H,G(t)]\p\qquad\p\in\Phi \]
The subalgebra $\G':=\{G_j\in\G|[G_j,H]=0\}$ is a more
narrow definition of symmetry. The {\bf maximal symmetry algebra of} $H$
is defined to be the subalgebra $\G''\subset\G'$ of time-independent
operators commuting with the Hamilton operator.\\
The Heisenberg equation has the solution $G(t)=\e^{itH}G(0)\e^{-itH}$
where the evolution operator $\e^{itH}$ clearly commutes with the energy
operator $H$. Hence the time-dependent dynamical Lie algebra $\G=\{G_j(t
)|j=1,\ldots n\}$ and the time-independent dynamical Lie algebra
$\{G_j(0)|j=1\ldots n\}$ are unitarily equivalent which allows us to
restrict ourselves in concrete problems ot the analysis of
time-independent dynamical Lie algebras.\\
For stationary solutions of the Schr\"odinger equation of the form
$\p(t)=\e^{-iEt}u$ we obtain the eigenvalue equation $Hu=Eu$. An
eigenspace of $H$ for a fixed value $E$ of the energy is already a
representation space of the maximal symmetry algebra $\G''$ of $H$.
Hence $\G''$ should be rather called "algebra of degeneracy of the
energy". In order to solve the quantum mechanical problem completely,
we still have to determine the spectrum of $H$.\\
As an example let us analyze the spectrum of the nonrelativistic
hydrogen atom. Rotational symmetry of the Hamilton operator, i.e.
$[H,L_i]=0$, $i=1,2,3$, suggests $\so(3)$ as symmetry algebra
leading to a $2l+1$-fold degeneracy for each energy level. However,
this is just the kinematical symmetry algebra of the hydrogen atom. It
turns out that each energy eigenvalue $E_n$ has multiplicity $n^2$
(neglecting spin) independent of the angular momentum quantum number
$l$. For a given principal quantum number $n$, the eigenspace ${\cal H}
_n$ of $E_n$ can be decomposed into $2l+1$-dimensional irreducible
representations ${\cal D}(l)$ of $\so(3)$:
\baro {\cal H}_n&=& \bigoplus_{l=0}^{n-1}{\cal D}(l)\\
      \dim{\cal H}_n&=& \sum_{l=0}^{n-1}(2l+1)=n^2 \earo
This additional degeneracy is surprising and can only be explained in
terms of a higher ("hidden") symmetry of the Hamilton operator. In fact,
Pauli showed that the classical Runge-Lenz vector which occurs as a
constant of motion in the classical Kepler problem, leads to three
hermitian quantum-mechanical operators commuting with the Hamilton
operator. We conclude that the maximal symmetry algebra of $H$ is the
Lie algebra $\so(4)$, i.e. for a given $n$, the eigenspace ${\cal H
}_n$ of $E_n<0$ (bound states) carries a single $n^2$-dimensional
irreducible representation of $\so(4)$ (For $E_n>0$, the continuum
states, we have $\so(3,1)$). Thus $\so(4)$ may be interpreted
as the degeneracy algebra of the nonrelativistic hydrogen atom.\\
If the states of the hydrogen atom are labeled by the traditional
quantum numbers $|nlm\rangle$ associated with the solution in spherical
coordinates then, in constructing the full dynamical algebra, we must
find an algebra which contains $\so(4)$ as a subalgebra and
includes operators that ladder $n$ and $l$. A careful analysis exhibits
the $15$-dimensional Lie algebra $\so(4,2)$ as a dynamical algebra
for the hydrogen atom which means that its operators permit us to pass
from any hydrogenic state $|nlm\rangle$ to any other state $|n'l'm'
\rangle$. Hence there is a {\it single} irreducible representation of
$\so(4,2)$ that covers {\it all} the states of the hydrogen atom.
Of course this representation must be infinite-dimensional.\\
It is worth mentioning that already $\so(4,1)$ possesses a single
irreducible representation which covers the complete set of quantum
numbers $n,l,m$, and may therefore be regarded as the quantum-number
algebra of the hydrogen atom. The enlargement of $\so(4,1)$ to
$\so(4,2)$ introduces no additional quantum nubers and leaves the
representation space unchanged but it requires additional operators
that can be identified with interaction operators. Thus, in principle,
the calculation of electromagnetic transition amplitudes and the Stark
effect has been reduced to an algebraic calculation without any need to
compute integrals. It is also remarkable that the generators of
$\so(4,2)$ may be realized in terms of the four-dimensional Dirac
$\gamma$-matrices.\\[1em]
After this digression on symmetries it is tempting to speculate about
applications of Borcherds algebras in physics. It might be possible to
construct quantum field theories in which a Borcherds algebra plays the
role of a sort of dynamical Lie algebra. One would expect to find all
quantum states within a single representation. In particular, the
dynamical algebra should comprise the Hamilton operator as well as
operators that change number of particles. The underlying Lie algebra
without Borcherds extension then could determine the maximal symmetry
algebra of the Hamilton operator.

Another area where Borcherds algebras might emerge is string field
theory. It is astonishing that the irreducible representations of the
above discussed examples precisely match the Fock space of bosonic
string field theory with the underlying Kac-Moody algebra as
spectrum-generating algebra for the 1-string Hilbert space.

After discussing the simplest examples of Borcherds algebras we finally
want to return to the probably least trivial example of a Borcherds
algebra, namely, the Monster Lie algebra invented by Borcherds in \cite%
{Borc92}.
\subsection {The Monster Lie algebra} \labelx{5.3}
In \cite{FLM88}, Frenkel, Lepowsky and Meurman constructed the Monster
vertex algebra which is acted on by the Monster simple sporadic group.
The underlying vector space which is called {\bf Moonshine Module}
\cite{FrLeMe85} and is denoted by $\F^\natural$ provides a natural
infinite-dimensional representation of the Monster is characterized by
the following properties: \beginnen
 \item[(i)] $\F^\natural$ is a vertex operator algebra with a conformal
            vector $\w$ of dimension 24 and a positive definite bilinear
            form
 \item[(ii)] $\F^\natural=\bigoplus_{n\ge-1}\F^\natural_n$ where
             $\F^\natural_n\equiv\Fo{n+1}^\natural$ is the eigenspace of
             $\Lo$ with eigenvalue $n+1$, and the dimension of
             $\F^\natural_n$ is given via the generating function
  \[ \sum_{n\ge-1}\dim\F^\natural_nq^n=J(q)\equiv j(q)-744
                                           =q^{-1}+196884q+\ldots \]
 \item[(iii)] The Monster group acts on $\F^\natural$ preserving the
              vertex operator algebra structure, the conformal vector
              $\w$ and the bilinear form \beenden
(Note the shift of weights!) It is crucial that the constant term of
$J(q)$ is zero which entails that there are no primary fields of weight
one. Remember that this property was essential for constructing the
Griess algebra of primary fields of weight two in Subsection \ref{3.5}.
Indeed, $\F^\natural_{1}$ coincides with the Griess algebra since
additionally the spectrum of $\Lo$ is nonegative and $\F^\natural_{-1}$
is one-dimensional.

The Monster vertex algebra is realized explicitly as
\[ \F^\natural=\F^+_\Leech\oplus\F'^+_\Leech \]
where $\F_\Leech$ denotes the vertex operator algebra associated with
the Leech lattice, the unique 24-dimensional even unimodular Euclidian
lattice with no elements of square length two. The symbol '+' denotes
the subspace fixed by a certain involution whereas the prime indicates
a twisted construction involving the square roots of the formal
variables. A readable account to the Monster module as $\Z_2$--orbifold
of a bosonic string theory compactified to the Leech lattice can be
found in \cite{Tuit92} whereas a $\Z_p$--orbifold description is
presented in \cite{DonMas2}. It is interesting that there is also an
approach to the Monster module based on twisting the heterotic string.
\cite{Harv85}

The starting point for the definition of a Monster Lie algebra should be
the fake Monster Lie algebra. We use the fact that the Lorentzian
lattice $\II$ can be written as the direct sum of the Leech lattice and
the unique two-dimensional even unimodular Lorentzian lattice $\III$.
The elements of $\III$ are usually represented as pairs $(m,n)\in\Z
\oplus\Z$ with inner product matrix
$\left({\ 0\atop-1}{-1\atop\ 0}\right)$
so that $(m,n)^2=-2mn$.
\setcounter{figure}{1}
\begin{figure}[htb] \unitlength.6cm \begin{center}
\begin{picture}(18,18)
\linethickness{.05mm}
\multiput(1,0)(2,0){4}{\line(0,1){18}}
\multiput(11,0)(2,0){4}{\line(0,1){18}}
\multiput(0,1)(0,2){4}{\line(1,0){18}}
\multiput(0,11)(0,2){4}{\line(1,0){18}}
\linethickness{.3mm}
\put(9,0){\vector(0,1){18}}   \put(9,18){\makebox(0,0)[b]{$x_2$}}
\put(0,9){\vector(1,0){18}}   \put(18,8.9){\makebox(0,0)[l]{$x_1$}}
\put(0,0){\vector(1,1){18}}   \put(18,18){\makebox(0,0)[lb]{$m$}}
\put(18,0){\vector(-1,1){18}} \put(0,18){\makebox(0,0)[rb]{$n$}}
\thinlines
\multiput(1,1)(4,0){5}{\begin{picture}(0,16) \multiput(0,0)(0,4){5}
                       {\circle*{ .3}} \end{picture}}
\multiput(2,1)(4,0){4}{\begin{picture}(2,16) \multiput(0,0)(1,1){3}
                       {\begin{picture}(0,16) \multiput(0,1)(0,4){4}
                       {\circle*{ .3}} \end{picture}} \end{picture}}
\put(1,8.9){\makebox(0,0)[tr]{\small-4}}
\put(3,8.9){\makebox(0,0)[tr]{\small-3}}
\put(5,8.9){\makebox(0,0)[tr]{\small-2}}
\put(7,8.9){\makebox(0,0)[tr]{\small-1}}
\put(9,8.9){\makebox(0,0)[tr]{\small0}}
\put(11,8.9){\makebox(0,0)[tr]{\small1}}
\put(13,8.9){\makebox(0,0)[tr]{\small2}}
\put(15,8.9){\makebox(0,0)[tr]{\small3}}
\put(17,8.9){\makebox(0,0)[tr]{\small4}}
\put(9,0.9){\makebox(0,0)[tr]{\small-4}}
\put(9,2.9){\makebox(0,0)[tr]{\small-3}}
\put(9,4.9){\makebox(0,0)[tr]{\small-2}}
\put(9,6.9){\makebox(0,0)[tr]{\small-1}}
\put(9,10.9){\makebox(0,0)[tr]{\small1}}
\put(9,12.9){\makebox(0,0)[tr]{\small2}}
\put(9,14.9){\makebox(0,0)[tr]{\small3}}
\put(9,16.9){\makebox(0,0)[tr]{\small4}}
\end{picture} \caption[lattice]{The Lorentzian lattice $\III$}
\end{center} \end{figure}
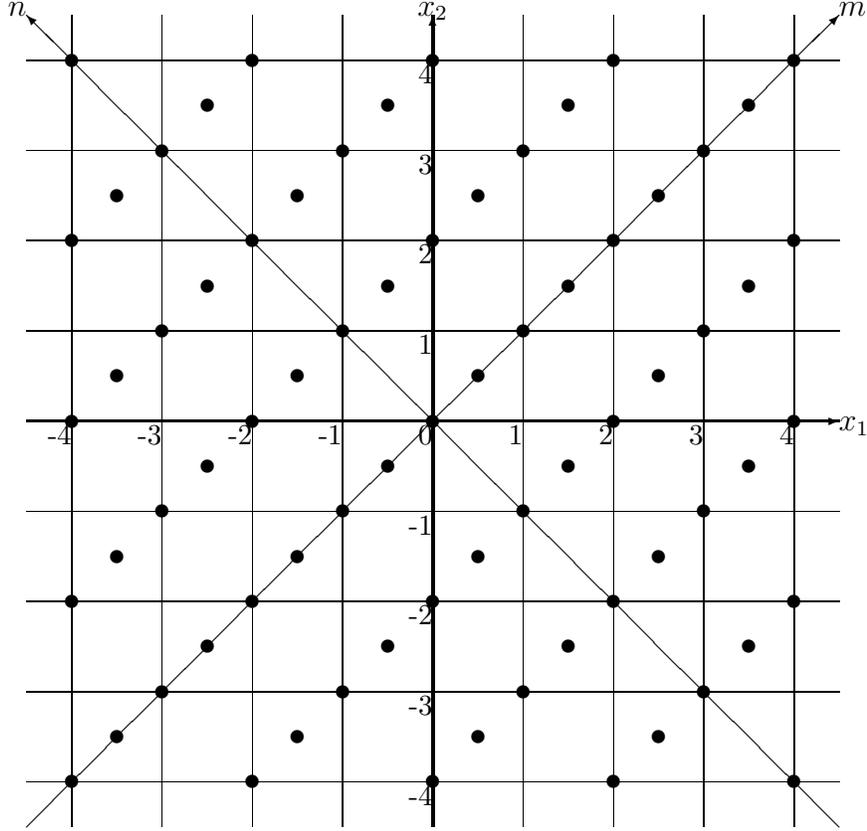

We need the following general result on vertex algebras \cite{FrHuLe93}.
Given two vertex algebras $(\F_i,$ $\V_i,$ $\1_i,$ $\w_i),$ $i=1,2$,
the vector space \[ \F:=\F_1\otimes\F_2 \] becomes a vertex algebra of
rank equal to the sum of the ranks of the $\F_i$ when we provide $\F$
with the tensor product grading and set
\baro \V(\p_1\otimes\p_2,z)&:=& \V_1(\p_1,z)\otimes\V_2(\p_2,z)
                               \for{}\p_i\in\F_i \\
      \1&:=&\1_1\otimes\1_2 \\
      \w&:=&\w_1\otimes\1_2+\1_1\otimes\w_2 \earo
In our case we obtain
\[ \F_{\II}=\F_{\Leech\oplus\III}=\F_{\Leech}\otimes\F_{\III} \]
and the vertex algebra associated with the Lorentzian lattice $\II$ is
the tensor product of the vertex algebras corresponding to $\F_{\Leech}$
and $\F_{\III}$. One finds that the Leech lattice gives rise to a vertex
operator algebra with conformal vector of dimension 24 and a positive
definite bilinear form. Furthermore, $\F_{\Leech}=\bigoplus_{n\ge-1}\F^{
\Leech}_n$ where $\F^{\Leech}_n\equiv\Fo{n+1}^{\Leech}$ is the
eigenspace of $\Lo$ with eigenvalue $n+1$, and the dimension of
$\F^{\Leech}_n$ is given via the generating function
\[ \sum_{n\ge-1}\dim\F^{\Leech}_nq^n=J(q)+24\equiv j(q)-720
                                           =q^{-1}+24+196884q+\ldots \]
The striking similarity between $\F_{\Leech}$ and the Moonshine module
$\F^\natural$ suggests the construction of the tensor vertex algebra
corresponding to $\F^\natural\otimes\F_{\III}$. Then the Monster Lie
algebra $\monster$ should be defined as the subspace
\[ \monster:=\Ph{1}^{\natural\otimes\III}\Big/{\rm kernel}(\_,\_)_
                       {\natural\otimes\III} \]
Obviously, $\monster$ is $\III$-graded, carries an invariant bilinear
form and has an involution which is induced by the trivial
automorphism of $\F^\natural$ and the natural involution $\theta$ of
$\F_{\III}$ (acting as $-1$ on $\III$ and on the piece of degree ${\bf0}
\in\III$ of $\F_{\III}$).

We can apply the no-ghost theorem of Goddard and Thorne in a modified
version of Borcherds to say more about the Monster Lie algebra.
\begin{theo} {\bf:}\labelx{t5} \\
Suppose that \beginnen
\item[(i)] $\F=\bigoplus_{n\ge-1}\F_n$ is a vertex operator algebra of
           rank 24 where $\F_n\equiv\Fo{n+1}$ is the piece of conformal
           weight $n+1$ (so that $\Lo$ has nonegative spectrum)
\item[(ii)] $\F$ is equipped with a nonsingular bilinear form $(\_,\_
            )_\F$ such that the adjoint of $\Ln$ is $\Lna$
\item[(iii)] $\F$ is acted on by a group $G$ which preserves all this
             structure  \beenden
Then we have the following natural $G$ module isomorphisms:
\baro \Ph{1),(\r}^{\otimes\III}\Big/{\rm kernel}(\_,\_)_{\F\otimes
      \F_{\III}}&\cong&\F_{-\frac12\r^2}\equiv\Fo{1-\frac12\r^2}\for{}
      {\bf0}\neq\r\in\III \\
    \Ph{1),({\bf0}}^{\otimes\III}\Big/{\rm kernel}(\_,\_)_{\F\otimes
      \F_{\III}}&\cong&\F_0\oplus\R^2\equiv\Fo{1}\oplus\R^2 \earo
where $\Ph{1),(\r}^{\otimes\III}$ denotes the subspace of degree $\r\in
\III$ of the physical space $\Ph{1}^{\otimes\III}=\{\p\in\F\otimes\F_
{\III}|\Ln\p=\delta_{n0}\p,n\ge0\}$, and $G$ acts trivially on $\F_{\III
}$ and $\R^2$. \end{theo}
Proof: See \cite{Borc92} \\[1em]
The no-ghost theorem implies that the piece of nonzero degree $(m,n)\in
\III$ of the Monster Lie algebra $\monster$ is isomorphic to the piece
$\F^\natural_{mn}$ of the Moonshine module and that the contravariant
form $(\_,\_)_0$ is positive definite on that piece of $\monster$. The
degree zero piece of $\monster$ is isomorphic to $\R^2$. Thus,
schematically the Monster Lie algebra looks like (cf. \cite{Borc92})
\[ \begin{array}{ccccccccccc}
   &\vdots&\vdots&\vdots&\vdots&{\textstyle n \atop \Big\uparrow}
                                      &\vdots&\vdots&\vdots&\vdots   \\
   \ldots& 0 & 0 & 0 & 0 & 0 &\Fn{4}&\Fn{8}&\Fn{12}&\Fn{16}&\ldots   \\
   \ldots& 0 & 0 & 0 & 0 & 0 &\Fn{3}&\Fn{6}&\Fn{9}&\Fn{12}&\ldots    \\
   \ldots& 0 & 0 & 0 & 0 & 0 &\Fn{2}&\Fn{4}&\Fn{6}&\Fn{8}&\ldots     \\
   \ldots& 0 & 0 & 0 &\Fn{-1}& 0 &\Fn{1}&\Fn{2}&\Fn{3}&\Fn{4}&\ldots \\
   -\hspace{-.5em}-\hspace{-.5em}-&0 & 0 & 0 & 0
                             &\R^2& 0 & 0 & 0 & 0 &\longrightarrow m \\
   \ldots&\Fn{4}&\Fn{3}&\Fn{2}&\Fn{1}& 0 &\Fn{-1}& 0 & 0 & 0 &\ldots \\
   \ldots&\Fn{8}&\Fn{6}&\Fn{4}&\Fn{2}& 0 & 0 & 0 & 0 & 0 &\ldots     \\
   \ldots&\Fn{12}&\Fn{9}&\Fn{6}&\Fn{3}& 0 & 0 & 0 & 0 & 0 &\ldots    \\
   \ldots&\Fn{16}&\Fn{12}&\Fn{8}&\Fn{4}& 0 & 0 & 0 & 0 & 0 &\ldots   \\
   &\vdots&\vdots&\vdots&\vdots&\Big|&\vdots&\vdots&\vdots&\vdots&
   \end{array} \]
If we finally define an element of $\monster$ of degree $(m,n)\in\III$
to have $\Z$-degree $2m+n$ then, with this $\Z$-grading, the Monster Lie
algebra is seen to be a generalized Kac-Moody algebra.

The $\III$-grading of $\monster$ looks like a root space decomposition
for the Monster Lie algebra and this suspicion turns out to be true. To
see this we go the other way round and try to find out what the
Borcherds algebra with root lattice $\III$ might be. Let us start with
the two vectors $\pm(1,-1)\in\III$ of square length two one of which
should be chosen as a real simple root, say, $(1,-1)$. Then we may take
$\Weyl:=(-1,0)$ as a (lightlike) Weyl vector so that the remaining
(imaginary!) simple roots $\r=(m,n)$ are determined by the condition
$\Weyl\cdot\r\stackrel!=-\frac12\r^2$ ($\iff n=mn$). Moreover we know
that simple roots must have non-positive inner product with each other.
We conclude that the vectors $(1,n)$, $n=-1$ or $n>0$, constitute a set
of simple roots for the root lattice $\III$.\\
The denominator formula for the corresponding Borcherds algebra then
reads
\[ p^{-1}\prod_{{m>0 \atop n\inZ}}(1-p^mq^n)^{{\rm mult}(m,n)}=
   \sum_{w\in{\cal W}}\det(w)w\left(p^{-1}\bigg(1-\sum_{n>0}{\rm mult}
   (1,n)pq^n\bigg)\right) \]
where we write $p$ and $q$ for the elements $\e^{(1,0)}$ and $\e^{(0,1)}
$ of the group ring of $\III$, respectively. Also note that all the
imaginary simple roots $(1,n)$, $n\ge1$, have nonzero inner product with
each other so that there is no extra contibution of pairwise orthogonal
imaginary simple roots on the right-hand side. Since the lattice $\III$
has only one real simple root the Weyl group has order two and is
generated by the corresponding reflection which exchanges $p$ and $q$
($\sigma_{(1,-1)}p=\e^{(1,0)-\{(1,-1)\cdot(1,0)\}(1,-1)}=q$). Hence
\baro p^{-1}\prod_{{m>0 \atop n\inZ}}(1-p^mq^n)^{{\rm mult}(m,n)}&=&
      p^{-1}-\sum_{n>0}{\rm mult}(1,n)q^n-
      \bigg(q^{-1}-\sum_{n>0}{\rm mult}(1,n)p^n\bigg) \\
      &=&\sum_{{n\ge-1 \atop n\neq0}}{\rm mult}(1,n)(p^n-q^n) \earo
where we used the fact that real simple roots always have multiplicity
one. Borcherds \cite{Borc92} was able to determine the unknown root
multiplicities by establishing an identity for the elliptic modular
function $j(q)$ which turned out to be precisely the above denominator
formula:
\[ p^{-1}\prod_{{m>0 \atop n\inZ}}(1-p^mq^n)^{c_{mn}}=j(p)-j(q) \]
with $j(q)-744=\sum_{n\ge-1}c_nq^n=q^{-1}+196884q+\ldots$

We summarize: The simple roots of the Monster Lie algebra $\monster$
are the vectors $(1,n)$, $n=-1$ or $n\ge1$, each with multiplicity
$c_n$.
\section{Omissions and Outlook}
In an introductory exposition of a rapidly developing area like vertex
algebras there are necessarily left out many interesting topics. In the
following we want to comment briefly on the omitted subjects of the
theory of vertex algebras as well as on the latest achievements.

First of all the formalism can be extended in order to include {\bf
twisted vertex operators} which involve square roots of the formal
variables. They are essential for constructing twisted repesentations of
affine Lie algebras. Moreover, as already mentioned, the construction
of the Moonshine module relies heavily on twisting vertex operators. For
this purpose the whole framework of formal calculus has to be slightly
modified such that it is also true for non-integral powers of the formal
variables \cite{FLM88}. In physics literature a similar treatment
can be found in \cite{DoGoMo90} and \cite{DoGoMo92}.

Another important issue is {\bf representation theory of vertex algebras
}. Categorical notions such as module, homomorphism, irreducibility,
simplicity etc. have to be defined properly in the new context. One
is led naturally to the definition of rational vertex operator algebras
which correspond to the familiar rational conformal field theories in
physics. The analogues of chiral vertex operators in conformal field
theory \cite{MooSei89} are precisely the intertwining operators for
vertex operator algebras introduced in \cite{FrHuLe93}. Therefore
the notion of fusion rules for vertex operator algebras also arises.
At present considerable effort is spent on representation theory of
vertex algebras associated with even lattices and in particular on the
study of the Moonshine module \cite{Dong1},\cite{Dong2},\cite{Dong3},
\cite{DonMas1}.

Recently, Schellekens \cite{Schel92} succeeded in classifying
$c=24$ self-dual chiral bosonic meromorphic conformal field theories
(see also \cite{Mont92}, \cite{Schel93}). His analysis was based upon
modular invariance of the partition function on the torus. Hence we can
ask how this result fits into the concept of vertex algebras. However,
as presented so far, there is no such thing like a modular parameter in
the framework of vertex algebras which just mirrows the fact that we
have dealt essentially with meromorphic conformal field theory on the
Riemann sphere. Three years ago Zhu \cite{Zhu90} introduced  {\bf
correlation functions on the torus} for vertex operator algebras and
found that, by taking graded dimensions of vertex operators of certain
vertex operator algebras, one obtains indeed modular forms. In particu%
lar, Zhu established modular invariance of the characters of the
irreducible representations as a general property of vertex operator
algebras. Translated into physics language this means that properties
of conformal field theory on the Riemann sphere directly imply the
corresponding properties on the torus.

Motivated by the question how to place the theory of Z-algebras and
parafermion algebras into an elegant axiomatic context and to embed
them into more natural algebras Dong and Lepowsky developed the theory
of vertex algebras much further. They defined generalized vertex
operator algebras and generalized vertex algebras and finally
introduced the even more general notion of {\bf abelian intertwining
algebras} where one-dimensional braid group representations are
incorporated intrinsically into the algebraic structure of vertex
algebras and their modules \cite{DonLep1}. This new avenue in the
formulation of vertex algebras is presented carefully in the monograph
\cite{DonLep2}.

Besides the ``standard'' approach to two-dimensional conformal quantum
field theory via states, fields, and operator products \cite{BePoZa84}
there is also a nice geometric formulation in terms of punctured Riemann
surfaces and sewing operations (\cite{Vafa87},\cite{GaGoMoVa88} or
\cite{West89} and references therein). In the
language of modular functors this was made mathematically rigorous by
Segal \cite{Sega88}, \cite{Sega89}. In the spirit of this geometric
concept Huang \cite{Huan90}, \cite{Huan91}, \cite{Huan92} introduced
{\bf geometric vertex operator algebras} whose essential ingredients
are the family of moduli spaces of $n$-punctured Riemann spheres and
the operations of sewing those spheres together. He was able to prove
the (categorical) equivalence of that definition with the familiar
notion of vertex operator algebras (Definition \ref{d2}) thereby
delivering an intrinsic geometric interpretation of vertex operator
algebras.

This relation was exploited even further in \cite{HuaLep93} where it
was argued that the moduli spaces of punctured Riemann spheres equipped
with the sewing operation constitute an algebraic structure which is
called analytic associative $\C^\times$--rescalable partial operad.
Thus vertex operator algebras may be reformulated in terms of {\bf
partial operads}. In this new language Stasheff \cite{Stas93} recently
also established connections to homotopy Lie algebras, Gerstenhaber
algebras \cite{PenSch92}, and Zwiebach's closed string field theory
\cite{Zwie92b},\cite{Zwie93}.

Motivated by recent progress in closed string field theory and in 2D
string theory Moore \cite{Moor93} undertook a renewed investigation
of the large symmetry algebras appearing in string theory. He considered
toroidal compactifications of all spacetime coordinates (cf. Section
\ref{4.1}!) and constructed new infinite-dimensional unbroken symmetry
algebras of the target space. In this formulation Borcherds algebras
emerge as examples of {\bf enhanced symmetry points} in the moduli space
of even unimodular lattices. In particular, the product of two copies
of the fake Monster Lie algebra can be regarded as a maximal symmetry
algebra of the theory since it corresponds to a unique point in the
Narain moduli space (of conformal field theories) at which the closed
bosonic string completely factorizes between left and right movers.

In his proof of Conway's and Norton's Moonshine conjectures for the
Moonshine module Borcherds \cite{Borc92} exhibited a whole family
of {\bf monstrous Lie superalgebras} similar to the Monster Lie algebra.
In fact, as the Monster Lie algebra corresponds to the identity element
of the Monster group, Borcherds defined generalized Kac-Moody super%
algebras associated with other elements of the Monster and worked out
their root multiplicities by using the twisted version of the denominat%
or formula (cf. Section \ref{5.1}).

These remarks and comments should be sufficient to show that vertex
algebras constitute a powerful framework in analyzing questions in
conformal quantum field theory and maybe closed string field theory.
It is a rapidly evolving area of mathematics which might provide us
with some deep insight into the foundations of physics. Moreover, Borch%
erds algebras when interpreted as generalized symmetry algebras,
seem to be the natural next step towards the formulation of a universal
symmetry in string theory.
\vspace{1cm}

I would like to thank all participants of the seminar on generalized
Kac-Moody algebras given at the II. Institute for Theoretical Physics
during spring '93 where I was allowed to present part of this material.
Their questions and suggestions were quite useful for compiling the
final version. I am especially grateful to Professor Slodowy for many
helpful comments and to Professor Nicolai for his constructive criticism
and for encouraging me to write this review.


\end{document}